\documentclass[twocolumn]{aastex63}
\pdfoutput=1 
\usepackage{amstext,amsmath}
\usepackage[T1]{fontenc}
\usepackage[figure,figure*]{hypcap}
\usepackage{comment}
\usepackage{physics}

\usepackage{amsmath}
\usepackage[bb=boondox]{mathalfa}
\usepackage{graphicx}
\graphicspath{ {./} }

\shorttitle{WDM Inference}
\shortauthors{Hibbard et al.}

\begin{document}

\title{Constraining Warm Dark Matter and Pop III stars with the Global 21-cm Signal}

\correspondingauthor{Joshua J. Hibbard}
\author{Joshua J. Hibbard}
\affiliation{Center for Astrophysics and Space Astronomy, Department of Astrophysical and Planetary Science, University of Colorado Boulder, CO 80309, USA}

\author{Jordan Mirocha}
\affiliation{Department of Physics \& McGill Space Institute, McGill University, 3600 Rue University, Montr\'eal, QC, H3A 2T8}

\author{David Rapetti}
\affiliation{NASA Ames Research Center, Moffett Field, CA 94035, USA}
\affiliation{Research Institute for Advanced Computer Science, Universities Space Research Association, Columbia, MD 21046, USA}
\affiliation{Center for Astrophysics and Space Astronomy, Department of Astrophysical and Planetary Science, University of Colorado Boulder, CO 80309, USA}

\author{Neil Bassett}
\affiliation{Center for Astrophysics and Space Astronomy, Department of Astrophysical and Planetary Science, University of Colorado Boulder, CO 80309, USA}

\author{Jack~O.~Burns}
\affiliation{Center for Astrophysics and Space Astronomy, Department of Astrophysical and Planetary Science, University of Colorado Boulder, CO 80309, USA}

\author{Keith Tauscher}
\affiliation{Center for Astrophysics and Space Astronomy, Department of Astrophysical and Planetary Science, University of Colorado Boulder, CO 80309, USA}
\affiliation{Department of Physics, University of Colorado, Boulder, CO 80309, USA}

\email{Joshua.Hibbard@colorado.edu}

\begin{abstract}
    Upcoming ground and space-based experiments may have sufficient accuracy to place significant constraints upon high-redshift star formation, Reionization, and dark matter (DM) using the global 21-cm signal of the intergalactic medium. In the early universe, when the relative abundance of low-mass DM halos is important, measuring the global signal would place constraints on the damping of structure formation caused by DM having a higher relic velocity (warm dark matter, or WDM) than in cold dark matter (CDM). Such damping, however, can be mimicked by altering the star formation efficiency (SFE) and difficult to detect because of the presence of Pop III stars with unknown properties. We study these various cases and their degeneracies with the WDM mass parameter $m_X$ using a Fisher matrix analysis. We study the $m_X = 7$ keV case and a star-formation model that parametrizes the SFE as a strong function of halo mass and include several variations of this model along with three different input noise levels for the likelihood; we also use a minimum halo virial temperature for collapse near the molecular cooling threshold. We find that when the likelihood includes only Pop II stars, $m_X$ is constrained to an uncertainty of $\sim 0.4$ keV for all models and noise levels at 68$\%$ CI. When the likelihood includes weak Pop III stars, $m_X \sim 0.3$ keV, and if Pop III star formation is relatively efficient, $m_X \sim 0.1$ keV uncertainty, with tight Pop III star-formation parameter constraints. Our results show that the global 21-cm signal is a promising test-bed for WDM models, even in the presence of strong degeneracies with astrophysical parameters.
    \end{abstract}


\section{Introduction}
\label{sec-WDM}
At this current juncture in astrophysics there are numerous models for DM, all of which are characterized by either their interaction cross-sections with the Standard Model particles, self-interactions (annihilation and decay), or impact on cosmological structure. WDM falls under the latter class of models, and is generally defined as DM that suppresses small-scale structure (i.e., dwarf galaxies) but still falls in line with current observational constraints \citep{Primack:1997,Weinberg:2015,Schneider:2015}. Possible Standard Model particle candidates for this kind of WDM include both the gravitino and the sterile, right-handed neutrino. The latter in particular is well-motivated, as it can solve the problem of the mixing angles of active neutrinos \citep{Drewes:2013}.

WDM is posited as a solution to several observational problems associated with small-scale structure--or rather, the lack thereof. Indeed, the \textit{missing satellite} \citep{Klypin:1999, Moore:1999b} and \textit{too big to fail} \citep{Boylan-Kolchin:2011} problems can theoretically be alleviated by a studious choice of WDM thermal mass. In the latter problem, DM halos which are too massive to not host more collapsing baryonic structure have been observed, and in the former there are far too few satellite (or dwarf) galaxies observed to be orbiting large galaxies such as the Milky Way. CDM simulations predict hundreds to thousands of satellite galaxies, depending on the size of the host galaxy or group, and merely dozens have been observed. Furthermore, numerical simulations of DM halos tend to produce central regions (in small halos) that exhibit steep density profiles, in conflict with observations which tend to give nearly constant DM density profiles (the \textit{core-cusp} problem) \citep{Flores:1994,NFW:1997,Moore:1999a}, and some models of WDM are capable of alleviating such a discrepancy \citep{Avila-Reese:2001}; however, the thermal mass required for WDM to resolve the latter problem is in high tension with other observational constraints, especially those imposed by the Lyman-$\alpha$ forest (see below). For a nice review of the observational problems with CDM and the potential alleviation provided by WDM, see \cite{Weinberg:2015}.

It is possible that many of these problems are related to poorly understood feedback mechanisms generated by baryonic physics, such as stellar feedback, winds, and/or reionization feedback \citep{Brooks:2014}. Such processes could work to eliminate or prevent collapse of baryonic matter in low-mass DM halos and thus render them invisible observationally. Indeed, the star formation efficiency (SFE) probably varies with halo mass \citep{Mason:2015,Mashian:2016,Sun:2016}, and if the SFE drops too rapidly as a function of decreasing halo mass, one might na\"{i}vely expect a turnover in the SFE at low-mass to be undetectable.

However, the physics of star formation is also expected to change in low-mass halos \citep{Abel:2002,Bromm:1999}, so a simple extrapolation of the low-mass SFE as inferred from rest ultra-violet luminosity functions (UVLFs) may dramatically under-predict the amount of star formation at high-z. In particular, at high-redshifts, the low-metallicity, low-mass halos can host Pop III star formation, the presence of which would effectively alter the shape of the SFE function at low-masses, and which are also expected to have an effect on the 21-cm background \citep{Qin2021,Munoz2021,Magg2021}. Thus, it is not just useful to understand how all these confounding effects might hamper constraints on structure formation and DM---warm or otherwise---it is \textit{essential} that we determine if we can learn anything about DM's characteristics and the scale of suppression in the presence of such challenges.

Independent structure formation measurements have put tight constraints upon the WDM mass. Counts of dwarf galaxies rule out WDM masses $m_X < 2.3 $ keV (\cite{Kennedy:2014}), measurements of UV luminosity functions of lensed galaxies in the Hubble Frontier Field at $z\approx6$ rule out masses below $m_X = 2.4$ keV (\cite{Menci:2016}), Lyman$-\alpha$ forest data (\cite{Viel:2013}) rule out WDM masses $< 3.3$ keV, and \textit{N}-body simulations of high redshift halo mass functions (\cite{Shirasaki:2021}) rule out WDM masses $< 2.71$ keV, all at the $95\%$ confidence level. Even more stringent, recent combinations of strong gravitational lensing of host halo masses with Milky Way satellite subhalo populations have disfavored $m_X < 9.7$ keV, also at the $95\%$ confidence level \citep{Nadler:2021}. Much work in WDM has also been done in the field of semi-analytic, high-redshift galaxy formation and evolution for masses $m_X < 5$ keV \citep{Dayal:2015,Dayal:2017}. 

Each of these observations has its own unique observational challenges and systematics to overcome, and all of them are highly model-dependent and locally calibrated. In fact, the majority of the WDM constraints cited concern small-scale structure in the local Universe ($z\lesssim6$). Moreover, as WDM is primarily a theory about the abundance and turnover of small-mass halos, there is perhaps nowhere its effects will be as readily apparent and testable as within the high redshift Universe, particularly the Cosmic Dawn and Epoch of Reionization \citep{Sitwell:2014,Lopez-Honorez:2017, Schneider:2018, Lopez-Honorez:2019, Munoz:2018, Yoshiura:2020}. Thus, an analysis of the effects of WDM upon the global 21-cm signal (henceforth, global signal) allows for the possibility of WDM mass constraints in epochs in which such measurements have never been performed and which entertain exquisite sensitivity to small-scale structure.Such 21-cm analyses can thus further help to constrain the small-scale matter power spectrum, including the effects of WDM, as seen in \cite{Munoz:2020}.

To date, no attempt has been made to carefully quantify the degeneracies between WDM and astrophysical parameters, nor has anyone (as far as the authors are aware) included multiple astrophysical parametrizations and stellar populations. In this paper we present the first such quantification using the Fisher information matrix, allowing us to determine the extent to which the parameters controlling the cosmology of the global signal (such as the WDM thermal mass) can be constrained when including the astrophysics of high-redshift star-formation. To be comprehensive, we examine two common star-formation parametrizations found in the literature, multiple input data noise covariances, two SFE models, and also include Pop II and Pop III stellar populations in our modeling. Our results also double as a forecast for Pop III parameters, which hasn't yet been extensively explored in the context of the global signal.

The rest of this paper is organized as follow: Section \ref{sec-methodology} describes our implementation of WDM into the standard global signal calculations, including several different star-formation parametrizations, stellar populations, and input noise covariances; Section \ref{sec-results} presents our Fisher forecasts which examine the astrophysical and cosmological parameter constraints; we include a brief discussion of other possible models which could be incorporated in Section \ref{sec-discussion}; and we conclude in Section \ref{sec-conclusions}.

\section{Methodology}
\label{sec-methodology}
\subsection{Warm Dark Matter}
The effects of WDM upon the global 21-cm signal can be incorporated by a suitable choice of the DM halo mass function (HMF), or the number of DM halos per unit mass per co-moving volume of the Universe, usually written as (e.g. in \cite{Press:1974})
\begin{equation}
    \frac{dn}{dM} = \frac{\Bar{\rho}}{M^2}  f(\sigma) \Bigg| \frac{d \ln{\sigma}}{d \ln{M}} \Bigg|.
    \label{eqn-hmf}
\end{equation}
Here, $n$ is the number of halos, $M$ is the mass of each halo, $\Bar{\rho}$ is the mean matter density of the Universe, $\sigma$ is the variance of the overdensity fluctuations, and $f(\sigma)$ is a fitting function that depends upon the model for halo collapse (see \cite{Binney:08}). In practice, we are interested in the variance of the density fluctuations smoothed by a window function. For WDM, a common choice for the latter is specifically the so-called \textit{sharp k-space} filter (see below for a discussion of this choice of filter model), which is a step function in Fourier amplitude k-space:
\begin{equation}
    \widetilde{W}_K(\vb k) \equiv \Theta(1 - k/K)
\end{equation}
where $\Theta(x)$ is the Heaviside step function and $K$ represents the characteristic scale of the window function. Thus, all fluctuations with amplitudes $k > K \sim R^{-1}$ are set to zero with this window function, where $R$ defines a given radius of the enclosed mass of the halo. Convolving the variance with this filter, we find the variance smoothed on a scale $K$ to be 
\begin{equation}
    \sigma_K^2 = \frac{1}{2 \pi^2} \int_0^K dk k^2 P_{lin}(k)
\end{equation}
where $P_{lin}(k)$ is the linear CDM power spectrum leftover from inflation.
 
WDM is defined by having a non-negligible thermal relic velocity at decoupling (while the Universe is still radiation-dominated), which changes, with respect to CDM, both the primordial (or inflationary) linear matter power spectrum and the nonlinear collapse of DM halos. In the case of the former, small-scale density fluctuations are suppressed due to the free streaming of WDM particles driven by their thermal velocities. This diffusion of small-scale density fluctuations erases the smaller DM halos which would have grown via gravitational collapse, and thus there is a characteristic cutoff length scale (or wave vector) in the linear matter power spectrum below which no DM halos form. This effect upon the linear matter power spectrum can be modeled with a transfer function fitted to numerical N-body simulations of WDM. A commonly used WDM linear power spectrum model was introduced by \cite{Bode:2001}:
\begin{equation}
    P_{X}(k) = T^2_{X}(k) P_{lin}(k)
\end{equation}
where $T_{X}(k)$ is the WDM transfer function, given by
\begin{equation}
    T_{X}(k) = (1 + (\alpha k)^{2 \nu} )^{-5/\nu}
\end{equation}
where $\nu = 1.2$, and
\begin{equation}
    \alpha = 0.048 \left(\frac{\Omega_{X0}}{0.4} \right)^{0.15} \left(\frac{h}{0.65} \right)^{1/3} \left(\frac{1}{m_{X}} \right)^{1.15} \left(\frac{1.5}{g_{X}} \right)^{0.29}.
\end{equation}
$\Omega_{X0}$ is the DM density parameter at the present time, $h$ is the reduced Hubble constant, $m_X$ is the WDM thermal mass in keV, and $g_X$ accounts for the abundance of WDM relative to photons and is set to a fiducial value of 1.5 for light neutrinos.

With this prescription, we can write the variance of the filtered WDM power spectrum $\sigma_X^2$ as 
\begin{equation}
    \sigma_X^2 = \frac{1}{2 \pi^2} \int_0^K dk k^2 T^2_X(k) P_{lin}(k).
\end{equation}

Nonlinear structure formation is well described by Extended Press-Schechter theory (EPS), which is, conveniently, analytical, and assumes that perturbations or fluctuations grow linearly with immediate, spherical halo formation above a certain threshold (see \cite{Press:1974}). The fitting function $f(\sigma)$ is derived from excursion-set theory by following the random-walk paths of overdensities as they diffuse towards the collapse threshold (\cite{Bond:1991}). In the spherical collapse case,
\begin{equation}
    f_{PS}(\sigma) = \sqrt{\frac{2}{\pi}} \frac{\delta_c}{\sigma} \exp{-\frac{\delta_c^2}{2 \sigma^2}},
\end{equation}
 where $\delta_c$ is the critical overdensity spherical collapse threshold, $\delta_c \approx 1.686$.
 
\cite{Sheth:2001} presented a fitting function derived from a model of ellipsoidal halo collapse, given by
\begin{equation}
    f_{ST}(\sigma) = A \sqrt{\frac{2a}{\pi}} \left[ 1 + \left( \frac{\sigma^2}{a \delta_c^2} \right)^p \right] \frac{\delta_c}{\sigma} \exp{\frac{-a \delta_c^2}{2 \sigma^2}},
    \label{eqn-ST-function}
\end{equation}
with parameters $A = 0.3222$, $a = 1$, and $p = 0.3$. Both collapse models agree reasonably with CDM simulations for masses $< 10^{13} h^{-1} M_{\odot}$, but tend to overpredict halo abundances for larger halo masses (\cite{Schneider:2015}). Because of this, \cite{Sheth:2001} allowed $a$ to be fit as a free parameter in their simulations and found better agreement for $a = 0.707$. This modified fitting function predicts the abundance of low and high mass CDM halos better than that assuming spherical collapse \citep[see for instance][]{White:2002, Lukic:2007, Binney:08}. However, both of these approaches fail disastrously to predict halo abundances when the cutoff scale is steeper than the typical CDM power spectrum, such as in the case of WDM. See \cite{Schneider:2013} and \cite{Benson:2013}, for some relevant discussions.

\cite{Schneider:2015}, however, found that fitting WDM simulations to the Sheth-Tormen HMF with $a = 1$ and using a sharp k-space filter to smooth the overdensity variance produced excellent agreement between numerical simulations of WDM and the analytical fitting functions. This is because, for any radius, the integral of $\sigma_X$ depends more strongly upon the power spectrum steepness than the shape of the filter. In this case, the mass of a halo can be defined via
\begin{equation}
    M \equiv \frac{4\pi }{3} \Bar{\rho} (cR)^3 \sim \Bar{\rho} K^{-3}
\end{equation}
where $c$ is a fitting parameter with a fiducial value of $c=2.5$. Thus, in order to agree with simulations, it is necessary in either the sharp k-space model or the original Sheth-Tormen model to introduce a free parameter. In this work, we shall adopt the sharp k-space filter model of \cite{Schneider:2015} with $a = 0.707$.

\begin{figure}
    \centering
    \includegraphics[width=0.46\textwidth]{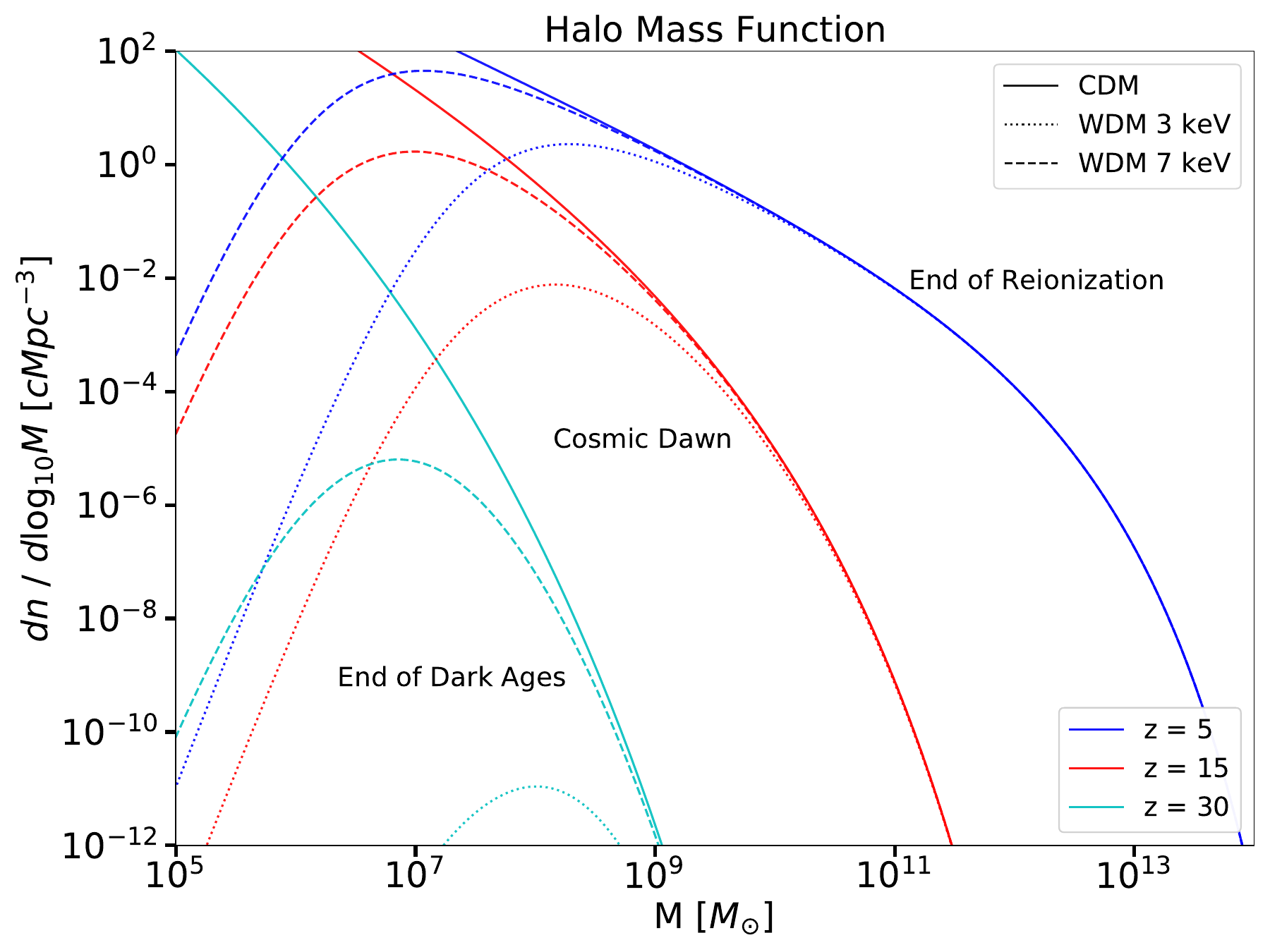}
    \caption{Halo Mass Functions for various DM models, assuming both a Sheth-Tormen fitting function and a sharp k-space window function.}
    \label{fig:hmf-various-dm-models}
\end{figure}

Figure \ref{fig:hmf-various-dm-models} shows the halo mass functions for CDM and two WDM cases, with $m_X = 3$ and $7$ keV, respectively, at redshifts 5, 15, and 30. These redshifts correspond roughly to the end of Reionization, Cosmic Dawn, and the end of the Dark Ages, respectively. We generate our halo mass functions using the python module \textsc{hmf}\footnote{https://github.com/halomod/hmf}, presented in \cite{Murray:2013}.

\subsection{Astrophysics and Star Formation Parametrization}
The simplest astrophysical parametrization of star formation assumes that the global rate of star-formation, or star-formation density (SFRD) is proportional to the rate of change in the collapsed mass density of the Universe, i.e. the mass in collapsed DM halos. However, as galaxies are treated in aggregate in these models, there is no distinction in star formation efficiency for different DM halo masses; that is, the efficiency with which baryons collapse into stellar, luminous material is independent of halo mass. Alternatively, one can construct star-formation models which still trace the collapse of halos, but have a star formation efficiency that is a function of the halo mass, peaking at the characteristic halo mass suitable for high-redshift galaxies. Both of these models are reviewed below.

\begin{figure*}
    \centering
    \includegraphics[width=0.96\textwidth]{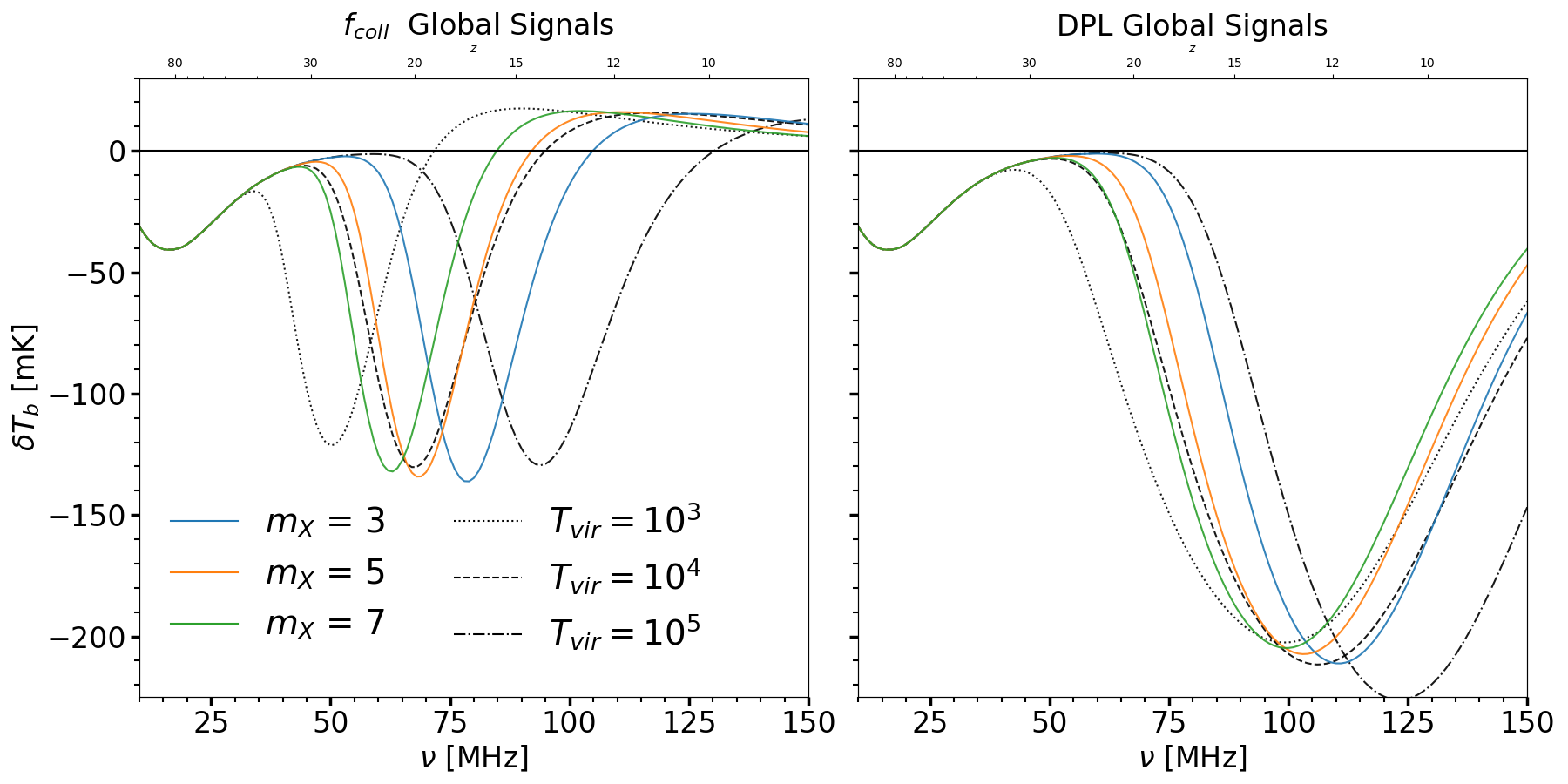}
    \caption{The global signal for various WDM model masses (blue, yellow, and green lines) and minimum CDM virial halo temperatures (dashed, dotted, and dashdot black lines) for two star formation models (see their descriptions in the text). The dashed black lines denote the atomic cooling threshold $T_{vir} = 10^4$ K. The trough near 10 MHz denotes the Dark Ages, while the second trough for each model denotes the Cosmic Dawn trough.}
    \label{fig:global-signal}
\end{figure*}

\subsubsection{Collapse Fraction star formation model}

In this classic star-formation rate parametrization \citep{Barkana:2005,Furlanetto:2006-nonreview}, the high-redshift galaxies are treated in aggregate, and their emissivity is proportional to the amount of DM which has collapsed into virialized halos. These models are also known as $f_{coll}$ models, where
\begin{equation}
    f_{coll}(z) = \Bar{\rho}_{m0}^{-1}(z) \int_{M_{min}}^{\infty} M \frac{dn}{dM} dM.
\end{equation}
In this equation $\Bar{\rho}_{m0}$ represents the mean co-moving mass density (baryons and DM) today, $dn / dM$ is the HMF from above (see Equation \ref{eqn-hmf}), and $M_{min}$ is the minimum threshold mass to collapse into a DM halo. This mass can be related to the virial temperature of the halo through the usual formula from the virial theorem for a monatomic gas obeying the ideal gas law:
\begin{equation}
    T_{vir} \equiv T_{min} = \frac{GM_{min} \mu m_{H}}{5R k_B}
\end{equation}
where $\mu$ is the mean molecular weight, $m_{H}$ the mass of Hydrogen, $G$ the universal gravitational constant, $k_B$ Boltzmann's constant, and $R$ is the (spherical) radius of the halo.

In general, this temperature is set by a particular cooling mechanism, such as $H_2$ cooling (in the case of Pop III stars) or simply atomic cooling of neutral Hydrogen for more massive halos. In our simulations, we set it to $T_{vir} = 300 K$, which is slightly below the $H_2$ cooling threshold (\cite{Tegmark:1997}). This choice is discussed in more detail below. 

In this simple model, the star-formation rate is assumed to trace the collapse of DM into halos (and thus of baryons into halos). This means that the the star-formation rate density (i.e. per co-moving volume) is proportional to the collapse fraction, to which in turn the total emissivity is proportional (see \cite{Pritchard:12}). The former can be written as
\begin{equation}
    \dot \rho_{\star}(z) = \Bar{\rho}_{b0} f_{\star} \frac{d f_{coll}}{dt},
\end{equation}
where $\Bar{\rho}_{b0}$ is the mean baryon density of the Universe today, and $f_{\star}$ is the fraction of baryons converted into stars, or star-formation efficiency. With these definitions, we can then write down the emissivity (in units of $photons/s/cMpc^{3}$) of the collapsed objects for the UV, Lyman-Werner, and X-ray bands, respectively (\cite{Mirocha:14}):
\begin{eqnarray}
    \epsilon_{UV}(z) = f_{esc} c_b N_{ion} \dot \rho_{\star}(z) \\
    \epsilon_{LW}(z) = c_b N_{LW} \dot \rho_{\star}(z) \\
    \epsilon_{X,src}(z) = f_X c_X \dot \rho_{\star}(z)
\end{eqnarray}
Here $f_{esc}$ is the fraction of ionizing photons which escape the collapsed halos; $N_{ion}$ is the number of ionizing photons (i.e. UV, Lyman-$\alpha$) produced per baryon of star-formation, with a fiducial value of $\approx 4000$ photons/baryon; $c_b$ is a conversion factor that sets the number of baryons per solar mass with a value of $c_b \sim 9.7\times10^{56}$ baryons/$M_{\odot}$; $N_{LW}$ is the number of photons emitted per baryon of star-formation in the Lyman-Werner band, with a fiducial value of $\approx 9690$ (see below for a discussion on this band); $c_X$ is the normalization factor in the well known local X-ray luminosity-SFR relation of \cite{Mineo:2012} with units of luminosity per SFR and a fiducial value of $c_X = 3 \times 10^{39}$; and finally $f_X$ parametrizes ignorance in the temporal evolution of the normalization factor $c_X$ for the high redshift Universe.

The ionizing photons are important for determining the ionization rate density, while the X-ray emissivity heats the IGM and determines its kinetic temperature $T_K$. The last radiation source of interest comes from the Lyman-Werner (LW) photons (11.2 - 13.6 eV) which affect the absorption bands of molecular Hydrogen ($H_2$) and can also source Lyman-$\alpha$ emission either directly or through cascades. These photons are important for determining the Pop III SFR and the meta-galactic background (see below for a discussion of these effects).

The main parameters of interest, and certainly those with the largest uncertainty in this aggregate model of star formation, are the star-formation efficiency $f_{\star}$, the escape fraction of UV photons $f_{esc}$, the high redshift $L_{X} / SFR$ relation parametrized by $f_{X}$, and the number of Lyman-Werner photons, $N_{LW}$. We fix $f_{\ast} = 0.1$, as it is perfectly degenerate with the other parameters in the $f_{coll}$ parametrization, and varying it would likely lead to numerical artifacts in the Fisher forecasts.

In principle, the minimum virial temperature of star-forming halos $T_{min}$ is also a free parameter with a large inherent uncertainty. Indeed, it is possible that the $M_{min}$ corresponding to the virial temperature is greater than the mass $M_{turn}$ at which the turnover in the HMF occurs due to WDM. If this were the case in reality, then the physics governing the value of $M_{min}$ would make the global signal indistinguishable from CDM, even if the actual HMF had a turnover induced by e.g. WDM.

The left panel of Figure \ref{fig:global-signal} shows several example $f_{coll}$ global signals for the frequency range associated with the Cosmic Dawn and the Epoch of Reionization. As seen in the Figure \ref{fig:global-signal}, the $f_{coll}$ models are very sensitive to $T_{min}$ as all halos form stars with the same efficiency. Even a small change in the minimum virial temperature can drastically change the timing of the signal. In the next star formation model, this dependence on $T_{min}$ is far weaker, as SFE is assumed to be a function of halo mass.

\subsubsection{Double-Power Law star formation model}
For the commonly used double-power law (hereafter DPL) parametrization of star formation, we follow the work of \cite{Mirocha:17}, where it is assumed that the star-formation efficiency $f_{\star}(M)$ depends upon halo mass $M$ as: 
\begin{equation}
    f_{\star}(M) = \frac{f_{\star,0}}{\left( \frac{M}{M_p} \right)^{\gamma_{lo}} + \left( \frac{M}{M_p} \right)^{\gamma_{hi}}}.
    \label{eqn-sfe-dpl}
\end{equation}
Here $f_{\star,0}$ is the star-formation efficiency at peak mass, $M_p$, and $\gamma_{lo}$ and $\gamma_{hi}$ are the power law indices at low and high halo masses, respectively. This allows us to write down the star-formation rate as
\begin{equation}
    \dot M_{\star}(M,z) = f_{\star}(M) \dot M_b(M,z)
    \label{eqn-sfr-dpl}
\end{equation}
where $\dot M_b$ is the baryonic mass accretion rate (MAR), which we derive directly from the HMF. By assuming that DM halos evolve at a fixed number density, we can abundance match across different redshifts, an approach which ensures self-consistency \citep{Furlanetto:2017}. While this technique works well for HMFs under CDM, we have found that DM scenarios with steep cutoffs (such as WDM) tend to produce non-monotonic behavior in the SFR for low-mass halos near the cutoff scale. Because of this, we use the MARs at all halo masses and redshifts from a CDM linear matter power spectrum to calculate the MARs of WDM halos. To our knowledge, the MARs of halos in WDM models have not been compared to CDM in detail via N-body simulations.

We then write the luminosity of a particular halo in a given band $\nu$ as
\begin{equation}
    L_{\nu}(M,z) = \dot M_{\star}(M,z) \mathcal{L}_{\nu}
    \label{eqn-lum-dpl}
\end{equation}
where $\dot M_{\star}$ is the star-formation rate and $\mathcal{L}_{\nu}$ is a factor which sets the luminosity per unit star-formation and depends upon the band. The primary ones of interest are the photon productions at 1600\AA, in the Lyman Continuum, and in the Lyman-Werner bands, represented by $\mathcal{L}_{1600}$, $\mathcal{L}_{LyC}$, $\mathcal{L}_{LW}$ respectively. For a given stellar metallicity, these three bands are computed with the BPASS version 1.0 spectral population synthesis models without binaries of \cite{Eldridge:2009}. For X-ray sources, $\mathcal{L}_X$ is modeled using the multi-colour disk (MCD) spectrum of \cite{Mitsuda:1984}, representative of HMXBs. This spectrum is normalized according to the observed $L_X-SFR$ relation, which includes as a free parameter $f_X$. The latter also affects the heating rate density used to calculate the kinetic temperature of the gas.

\begin{figure}
    \centering
    \includegraphics[width=0.46\textwidth]{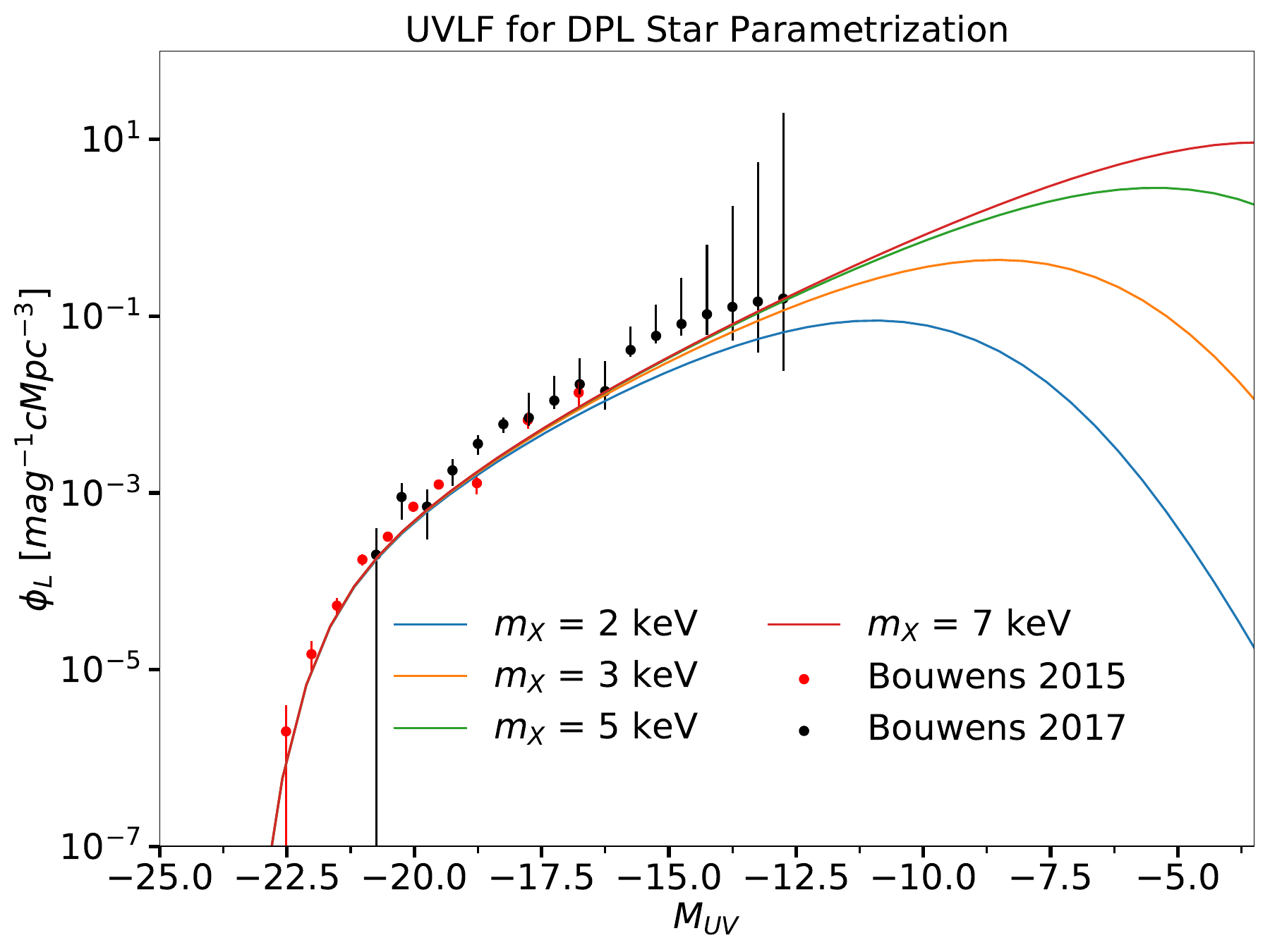}
    \caption{The rest-frame UV Luminosity Functions at $z \sim 6$ are plotted here for several different values of WDM mass $m_X$ (solid lines). The red and black dotted lines show the observational data of \citep{Bouwens:2015a, Bouwens:2017}, the former of which is used to calibrate the SFE parameters in the DPL parametrization. As can be seen, the high-redshift UVLFs alone disfavour WDM with $m_X \lesssim 2$ keV. Note that the $m_X = 7$ keV UVLF is the same one used to generate the DPL mock signals used in the DPL Fisher forecasts.}
    \label{fig:UVLF}
\end{figure}
Using Equations \ref{eqn-sfr-dpl}-\ref{eqn-lum-dpl}, we can write the intrinsic (rest-frame) galaxy UV luminosity function as follows:
\begin{equation}
    d\phi(L) = \frac{dn(M,z)}{dM} \left( \frac{dL}{dM} \right)^{-1} dL
    \label{eqn-lum-func}
\end{equation}
where $L$ is the observed UV luminosity, $dn/dM$ the halo mass function, and $\phi$ the galaxy luminosity function. Several example UVLFs for various WDM masses are plotted in Figure \ref{fig:UVLF}, including the $m_X$ = 7 keV UVLF which was used to generate the mock signals for the DPL Fisher forecasts (see below).

Finally, we can write the emissivity in each band $\epsilon_{\nu}$ for this star-formation model as
\begin{equation}
    \epsilon_{\nu}(z) = \int^{\infty}_{L_{min}} f_{esc, \nu} L_{\nu} \frac{d \phi(L_{\nu}, z)}{dL_{\nu}} dL_{\nu}
\end{equation}
where $f_{esc,\nu}$ parametrizes the escape fraction of photons for each band. We set $f_{esc,LW}$ to unity, a good approximation in the far-field limit \citep{Schauer:2017}, and leave $f_{esc,UV}$ to vary as a free parameter. Thus, WDM enters into this picture through the HMF and its effects upon the intrinsic luminosity function (Equation \ref{eqn-lum-func}), which ultimately characterizes the global emissivity.

The parameters of interest for the Fisher Matrix forecasts below using this star-formation model are the four DPL Pop II parameters $f_{\ast,0}, M_p, \gamma_{lo}$, $\gamma_{hi}$, the normalization of the $L_X/SFR$ relation, and $f_{esc, UV}$.

In the right panel of Figure \ref{fig:global-signal}, we plot several example global signals for the DPL star-formation model. As seen in the figure, these models are far less sensitive to the value of $T_{min}$, both in terms of its effects upon the timing and amplitude of the global signal. In all of the analysis that follows, we allow WDM to set the scale of suppression and fix $T_{min}$ to a constant, small value.

Although the DPL model is more complex than the $f_{coll}$ model, we can use published high-redshift galaxy rest ultra-violet luminosity functions (UVLFs) to constrain the SFE parameters. We use the UVLFs presented in \citet{Bouwens:2015a}, which are in good agreement with other studies \citep[see, e.g.,][]{Finkelstein2015}. We now examine how appending these UVLF data to the global signal likelihood can help constrain the DPL parameters.

\subsubsection{Jointly Fitting High Redshift UV-Luminosity Functions}
While there remains a paucity of data on UVLFs during and immediately after the Cosmic Dawn, we can in principle use the same UVLFs at a redshift of $z \sim 6$ which were used to calibrate the SFE of the DPL model in \cite{Mirocha:17} to help place constraints on the parameters appearing in our Fisher forecast. The data of \cite{Bouwens:2015a}, denoted as $\vb y_{UVLF}$ and plotted as red dots in Figure \ref{fig:UVLF}, can easily be appended to our likelihood, where the full data vector and model are now $\vb y = \begin{bmatrix} \vb y_{21} \\ \vb y_{UVLF} \end{bmatrix}$ and $\boldsymbol{\mathcal{M}} = \begin{bmatrix} \boldsymbol{\mathcal{M}}_{21}(\boldsymbol{\theta}) \\ \boldsymbol{\mathcal{M}}_{UVLF}(\boldsymbol{\theta}) \end{bmatrix}$, respectively. Note that the model for the signal and UVLF both take the same parameter vector, $\boldsymbol{\theta}$.

To finish up the DPL parametrization, we incorporate two additional models with parameters that can also change the timing of the signal, in order to exhaustively study other models with parameters which are degenerate with WDM: first, we introduce a simple extension to the DPL model to allow for a more flexible SFE, and finally we include the effects of Pop III stars.

\subsubsection{Extended Double-Power Law}

This model allows the SFE to depart from the DPL functional form of Equation \ref{eqn-sfe-dpl} in order to accommodate the possibility of unknown (i.e. un-modeled) physical processes or stellar populations at high redshifts.  We multiply Equation \ref{eqn-sfe-dpl} by a modulation factor of the form
\begin{equation}
    f_{\star} \rightarrow f_{\star} \left[ 1 + \left( \frac{M}{a} \right)^b \right]^c,
\end{equation}
where $a,b,c$ are new "nuisance" parameters which allow the SFE to depart from a power-law at the low mass end, degenerate, then, with the effects of adding WDM. This "extended double power-law" model, which we abbreviate as DPL XT, is employed in both \cite{Schneider:2021} and \cite{Mirocha:2021}, where it is pointed out that this parametrization allows for two different effects upon the low-mass halo population: (i) rapid decline in halos below a mass of $a$, mimicking either feedback processes which can inhibit star formation in low-mass halos or the effects of HMF cutoff scales introduced by, for example, WDM; (ii) it can boost the SFE $f_{\star}$ above those predicted by extrapolating UVLFs, which could in principle indicate the presence of Pop III stars; moreover, we also include a more physically-motivated treatment of Pop III stars, which we describe next.

\subsubsection{Pop III Stars}
\label{sec-popIIIstars-methods}
As Pop III stars may change the timing of the signal by introducing light before the nominal Cosmic Dawn of the Pop II stars, their effects are degenerate with those of WDM -- hence why we include them in our Fisher forecasts.

Our inclusion of Pop III stars follows \cite{Mirocha:2018}, which is summarized briefly here. Pop III stars ignite in metal-poor environments in halo masses below the virial temperature threshold for Pop II stars; we guess the initial Pop III virial halo temperature to be $\sim 500$ K, corresponding to the $H_2$ cooling threshold \citep{Tegmark:1997}, though this initial value is subsequently modified by LW feedback (see below). Additionally, we set the typical mass of our Pop III stars to be $\sim 100 M_{\odot}$, and calculate the SED (spectral energy distribution) for these high-mass, metal-poor Pop III stars from \cite{Schaerer:2002}, which give LyC and LW photons at rates of $\sim 1.4 \times 10^{50}$ and $1.6 \times 10^{50}$ photons per second over their lifetime, several orders of magnitude larger than Pop II cases. As LW photons can dissociate $H_2$ bonds and thus disrupt the cooling mechanism for forming Pop III stars in halos \citep{Haiman1997,Machacek2001}, we must solve iteratively for the meta-galactic radiation background produced by Pop~II and Pop~III sources and their subsequent effect upon the DM halo population (i.e. upon $T_{min}$ after each iteration). We relate the mean LW background intensity to the minimum mass of halos hosting Pop~III star formation via the fitting formulae in \cite{Visbal2014}.

We further assume that Pop III stars produce X-rays in a manner similar to Pop II stars, albeit with their own unique normalization factor to the $L_X/SFR$ relation, denoted as $(L_X/SFR)_{III}$. The overall amount of Pop~III star formation is  set by two parameters: the star formation rate in Pop~III halos, $\rm{SFR}_{\textsc{iii}}$, and the critical binding energy limit $BL$ that halos must exceed to transition from Pop III to Pop II star formation. Large values of $BL$ indicate strong Pop III star formation, as there will be many halos below this threshold that continue Pop III star-formation, due to their continued low metallicity (which is assumed to be due to the fact that energetic supernovae eliminate metals from the halo efficiently, due to the halo's low binding energy limit). Though this is a simple, phenomenological approach to Pop~III star formation, it is able to capture a broad range of possibilities that emerge in more physically-motivated semi-analytic models \citep[e.g.,][]{Trenti2009,Crosby2013,Mebane2018,Visbal2018}. 

Pop III stars can be thus be characterized by their X-ray radiation yield $(L_X/SFR)_{III}$, the binding energy limit $BL$ (measured in \textit{ergs}), and their star-formation rate $SFR_{III}$, which is assumed to be a constant.

\subsection{Fisher Matrix Parameter Forecasts}
We investigate the covariances between the astrophysical parameters and the WDM mass parameter by forecasting the Fisher Information Matrix (FIM) for each star formation model. Our Gaussian-distributed log-likelihood is given by
\begin{equation}
    \ln \mathcal{L}( \vb y| \boldsymbol{\theta}) \propto \left\{ -\frac{1}{2} [\vb y -  \boldsymbol{\mathcal{M}}( \boldsymbol{\theta})]^T \vb C^{-1} [\vb y - \boldsymbol{\mathcal{M}}(\boldsymbol{\theta)}] \right\}.
    \hspace{0.5cm}
\end{equation}

The global 21-cm signal data vector is $\boldsymbol{y}$ with dimension $n_y$ equal to the number of channels, the model as generated by a modified version of the publicly available code \textit{Accelerated Reionization Era Simulations} \textbf{ARES}\footnote{https://github.com/mirochaj/ares} \citep{Mirocha:12,Mirocha:14} for parameter vector $\boldsymbol{\theta}$ is $\boldsymbol{\mathcal{M}}$, and the input noise channel covariance is $\boldsymbol{C}$ with dimension $n_y \times n_y$. We will examine this likelihood with several different input noise covariances and study their effects upon parameter inference.

Our estimate of the FIM can then be written as
\begin{equation}
    \hat{F}_{ij} = - \frac{\partial^2 \ln \mathcal{L}(\boldsymbol{y}| \boldsymbol{\theta})}{\partial \theta_i \partial \theta_j} \Bigg|_{\boldsymbol{\theta} = \boldsymbol{\theta}_{MLE}}
    \label{eqn-fisher-matrix}
\end{equation}
where the maximum likelihood parameter vector $\boldsymbol{\theta}_{MLE}$ is given by the input values for each parameter in each star-formation model. The parameter covariance is then the inverse of the estimated FIM. For both star-formation models, we find that the Fisher matrix is stable over several orders of magnitude, with finite differences of about $\Delta \boldsymbol{\theta}_{j} \approx 0.001 - 0.1$.

\subsection{Likelihood Input Covariances}
\begin{figure}
    \centering
    \includegraphics[width=0.46\textwidth]{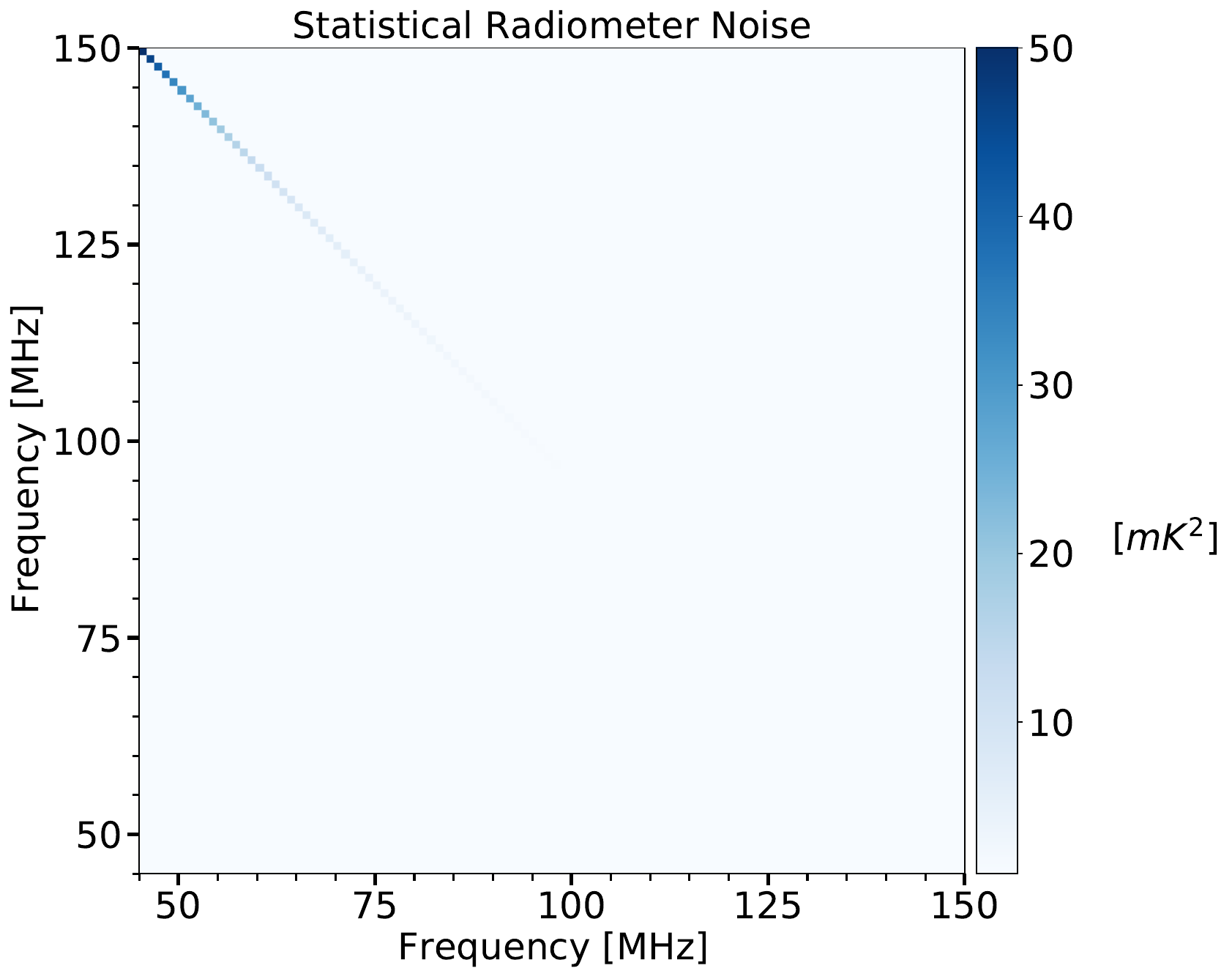}
    \includegraphics[width=0.46\textwidth]{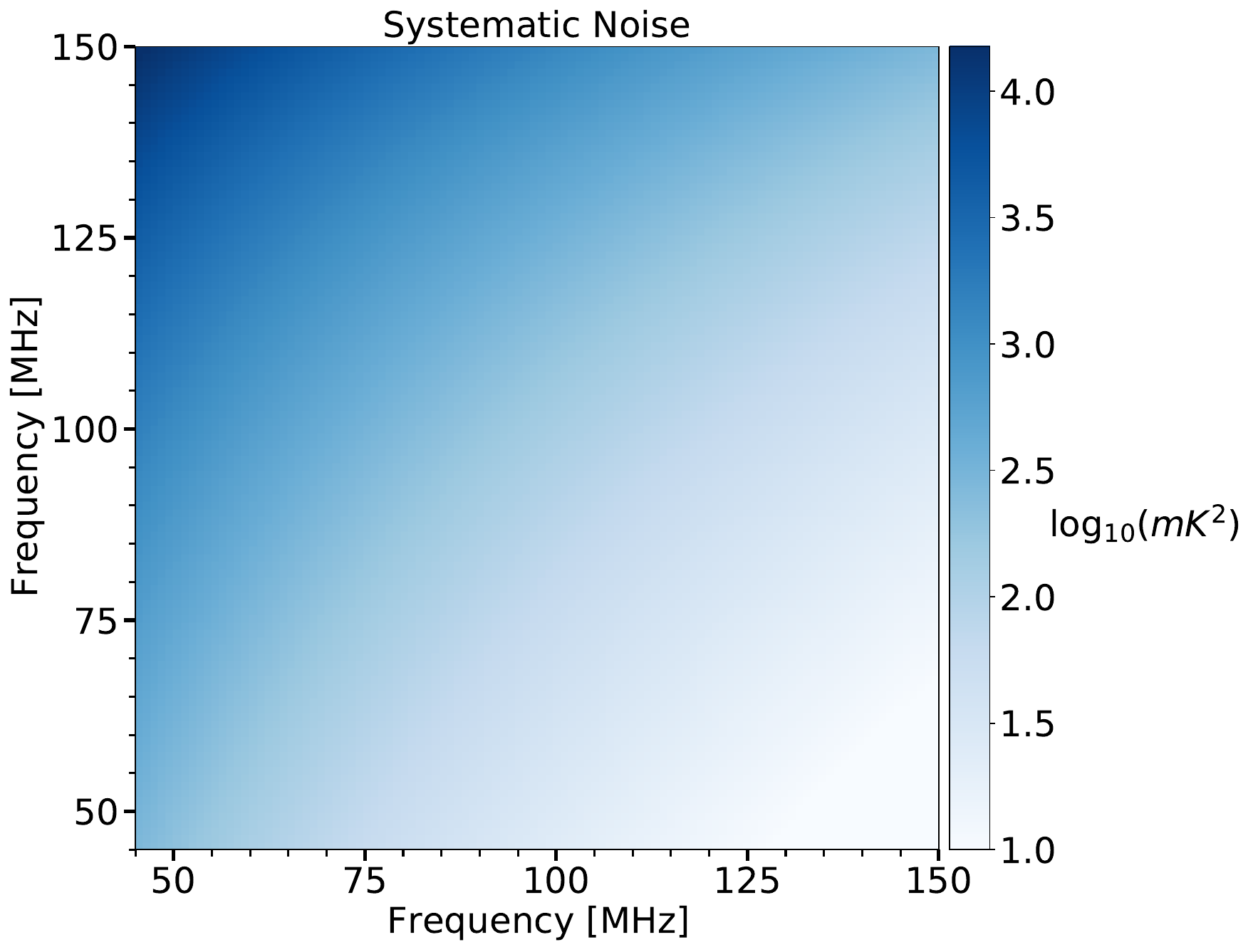}
    \includegraphics[width=0.46\textwidth]{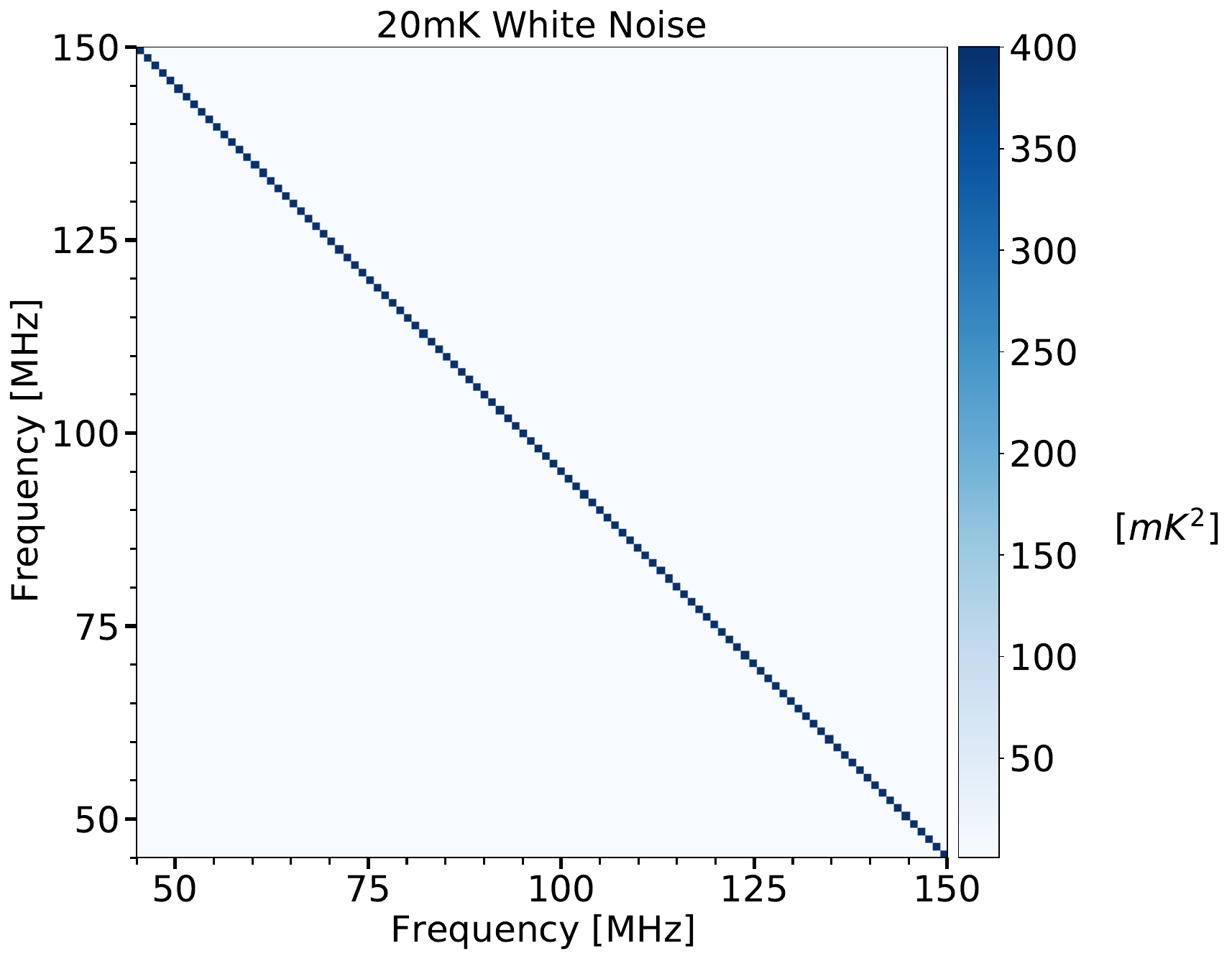}
    \caption{Input noise covariances for the likelihood and Fisher forecasts in the 45 - 150 MHz band. Both the Statistical Radiometer and Systematic noises are calculated using 250 hours of integration time for the simulated observation. \textit{Top Panel}: Statistical Radiometer noise is the ideal case, and corresponds to a diagonal, minimum error level with a low-band magnitude of $\sim 7$ mK error. \textit{Middle Panel}: Systematic noise has the same underlying inherent noise level in the parameters as the Statistical Radiometer case, but with the systematic noise effects of a beam and foreground included. Note that this plot is given in $\log_{10}$-space. \textit{Bottom Panel}: White noise is spectrally flat, diagonal, and with an error magnitude of 20 mK.}
    \label{fig:input-covariances}
\end{figure}

When sampling likelihoods to map out model parameter covariances, either in the case of a Fisher projection here or in an MCMC, an input noise covariance $\vb{C}$ is needed. We explore three different possibilities for this input covariance. For our simplest analyses, we adopt spectrally flat errors with a magnitude of $\sim 20$ mK, motivated by the noise level of the absorption trough reported by the EDGES experiment (\cite{Bowman:18}). We denote this spectrally flat noise as "White Noise," and display its covariance $\vb C$ in the bottom panel of Figure~\ref{fig:input-covariances}. 

More generally, a total-power receiver\footnote{A typical total-power receiver has thermal noise which is limited by the temperature of the \textit{receiver}. However, here we are assuming measurements in which the sky-noise (i.e. the foreground) dominates.} should produce noise which follows the ideal radiometer equation $\sigma_T \approx T_s / \sqrt{\Delta \nu \tau}$, where $T_s$ is the temperature of the sky, $\Delta \nu$ is the channel width, and $\tau$ is the total integration time in seconds (see \cite{Condon:16}). In our simulations, we take $\tau = 250$ hrs and $\Delta \nu = 1$ MHz. 

Because the low-frequency radio foreground follows mainly a power-law in the bandwidth of interest (\cite{Shaver:1999}, \cite{Furlanetto:06}), the noise on a total-power receiver is also expected to exhibit a spectral power-law form, determined by $T_s$ (discussed more below). In this work, we denote this type of noise as "Statistical Radiometer Noise," as it is the noise level we would expect in the absence of all confounding systematics and model overlap produced by beams and foregrounds; hence, it is the ideal case. Its covariance is shown in the top panel of Figure \ref{fig:input-covariances}, where its amplitude peaks at $\sim 7$ mK around 45 MHz. Ultimately, the value of $T_s$ used in the Statistical Radiometer Noise comes from the average spectrum of the realistic, beam-weighted foreground training set generated below; the same training set is also used to produce the Systematic Noise, to be discussed momentarily. Thus, the Statistical Radiometer and Systematic Noise cases have the same beam-weighted foreground model dominating the antenna noise, but the latter includes the \textit{additional} uncertainties that arise due to overlap between the global signal model and the beam-weighted foreground model, while the former does not.

The middle panel of Figure \ref{fig:input-covariances} shows the input noise covariance for "Systematic Noise." This is the noise covariance produced when including the contamination effects (specifically, the \textit{model overlap} between a global signal model and a beam-weighted foreground model) of the beam-weighted foreground. In order to produce this covariance, we first generate a foreground training set by starting with the 408 MHz all-sky temperature map of \cite{Haslam:82}, \cite{Remazeilles:2015} and then interpolating it down to 10 MHz assuming a spatially uniform spectral index of $\beta = -2.5$. Next, we assume a ten percent error in the temperature of each pixel in order to seed a uniform temperature distribution at this frequency; this distribution is then transformed \textit{back} into a spectral index distribution. This allows us to specify a temperature error per pixel from which we then generate a spectral index distribution, giving a whole-sky spectral index map seeded by errors in the sky temperature. This whole-sky spectral index map can then be used to extrapolate the Haslam map at 408 MHz down to low frequencies in the usual manner. This process can be run many times, producing a training set of unweighted foreground maps.

These maps are then convolved with a beam which is spatially Gaussian, with chromaticity that varies quadratically over the band of interest, identical to the ones used in \cite{Hibbard:2020}\footnote{This beam represents the ideal case; in practice, most beams will have large sidelobes which will change the measured temperature of the sky. Such a beam can be incorporated into our pipeline in the same manner, and would only slightly change the results of the Fisher forecast for the case of the Systematic noise. See \cite{Hibbard:2020} for examples of analyses which include beams with sidelobes.}. We split the beam-weighted convolutions of the training set into 10 equal LST bins, with our Gaussian antenna situated at the EDGES observatory latitude and longitude. We take the integration time for this simulated observation to be $\tau = 250$ hours with 1 MHz channel widths, as stated above. We generate basis vectors (or modes) for this training set using Singular Value Decomposition, as outlined in \cite{Tauscher:18}, assuming the Minimum Assumption Analysis (MAA) signal model with foreground priors taken from the foreground training set. The posterior channel (and, because it is the MAA, the parameter) covariance for a joint linear fit to the beam-weighted foreground SVD vectors and the MAA signal model is plotted in the middle panel of Figure \ref{fig:input-covariances}. As one can see, incorporating the beam-weighted foreground systematic overlap into the input noise covariance induces correlations between channels not seen in the case of the other input noise covariances; it is a non-diagonal matrix, and its elements are several orders of magnitude larger than in the case of the Statistical Radiometer Noise or the 20 mK White Noise.

\section{Results}
\label{sec-results}

\begin{table}
    \caption{Collapse Fraction SFR Model}
    \centering
    \begin{tabular}{c c c}
        \hline \hline
        Parameter & Input Value (MLE) & Prior \\
        \hline
        $f_{\star}$ & 0.1 (fixed) & Unif($-2.5, 0$)* \\
        $f_{X}$ & 1.0 & Unif($-1.3, 1.3$)* \\
        $N_{LW}$ & 9690 & Unif($2.67, 5.45$)* \\
        $f_{esc}$ & 0.1 & Unif($-2.75, 0$)* \\
        $m_X$ & 3, 7 & Unif(2, 10) \\
        \hline
        \multicolumn{3}{c}{* Parameter prior is shown in $\log_{10}$ space.}\\
        \hline
    \end{tabular}
    \label{tab:fcoll}
\end{table}

\begin{table}
    \caption{DPL SFR Model}
    \centering
    \begin{tabular}{c c c}
        \hline \hline
        Parameter & Input Value (MLE) & Prior \\
        \hline
        $f_{\star,0}$ & 0.05 & Unif($-3.5, 0$)* \\
        $M_p$ & $2.8 \times 10^{11}$ & Unif($11.0, 11.9$)* \\
        $\gamma_{lo}$ & 0.49 & Unif(0.3, 0.7) \\
        $\gamma_{hi}$ & -0.61 & Unif(-1.1, -0.1) \\
        $L_{X}/$SFR & $2.6 \times 10^{39}$ & Unif($39.25, 39.55$)* \\
        $f_{esc}$ & 0.2 & Unif($-0.95, -0.35$)* \\
        $m_X$ & 7 & Unif(2, 10) \\
        \hline
        \multicolumn{3}{c}{* Parameter prior is shown in $\log_{10}$ space.}\\
        \hline
    \end{tabular}
    \label{tab:dblpowerlaw}
\end{table}

\begin{table}
    \caption{Pop III Star Parameters}
    \centering
    \begin{tabular}{c c c}
        \hline \hline
        Parameter & Input Value (MLE) & Prior \\
        \hline
        $(L_X/SFR)_{III}$ & 2.6 $\times 10^{39}$ & Unif($36, 42$)* \\
        $BL$ & $5 \times 10^{51}$ & Unif($51.5, 52.1$)* \\
        $SFR_{III}$ & $5 \times 10^{-5}$ & Unif($-5, -2.5$)* \\
        \hline
        \multicolumn{3}{c}{* Parameter prior is shown in $\log_{10}$ space.}\\
        \hline
    \end{tabular}
    \label{tab:popIII}
\end{table}

\begin{table}
    \caption{Efficient Pop III Star Parameters}
    \centering
    \begin{tabular}{c c c}
        \hline \hline
        Parameter & Input Value (MLE) & Prior \\
        \hline
        $(L_X/SFR)_{III}$ & 2.6 $\times 10^{39}$ & Unif($36, 42$)* \\
        $BL$ & $1 \times 10^{52}$ & Unif($51.5, 52.1$)* \\
        $SFR_{III}$ & $1 \times 10^{-3}$ & Unif($-5, -2.5$)* \\
        \hline
        \multicolumn{3}{c}{* Parameter prior is shown in $\log_{10}$ space.}\\
        \hline
    \end{tabular}
    \label{tab:eff-popIII}
\end{table}

\subsection{Parameter Covariance: Collapse Fraction model}
\begin{figure*}
    \centering
    \includegraphics[width=0.96\textwidth]{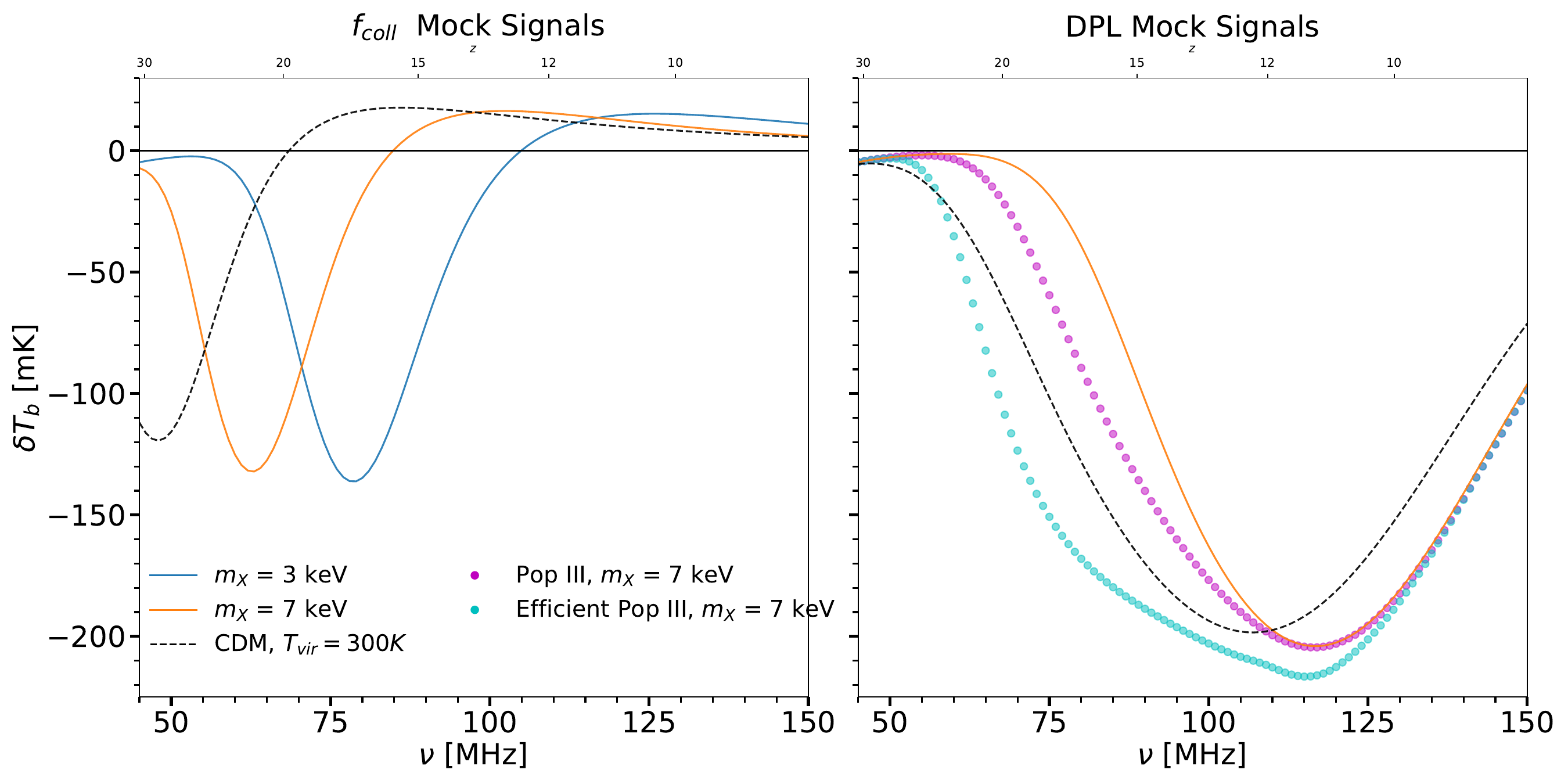}
    \caption{Input mock signals for both star-formation parametrizations, $f_{coll}$ and DPL, respectively. Note that the orange line for $m_X = 7$ keV labels the corresponding global signal for both star-formation parametrizations. The dotted magenta line indicates the Pop III mock signal case in which the Pop III SFE is low and halos transition quickly to Pop II star formation, hence the mild effects on the global signal. Additionally, a mock signal with efficient Pop III star formation is shown by the dotted cyan line. We show the equivalent CDM case for each parametrization with $T_{vir} = 300 K$ as a black, dashed line for comparison.}
    \label{fig:mock-signals}
\end{figure*}

\begin{figure*}
    \centering
    \includegraphics[width=0.96\textwidth]{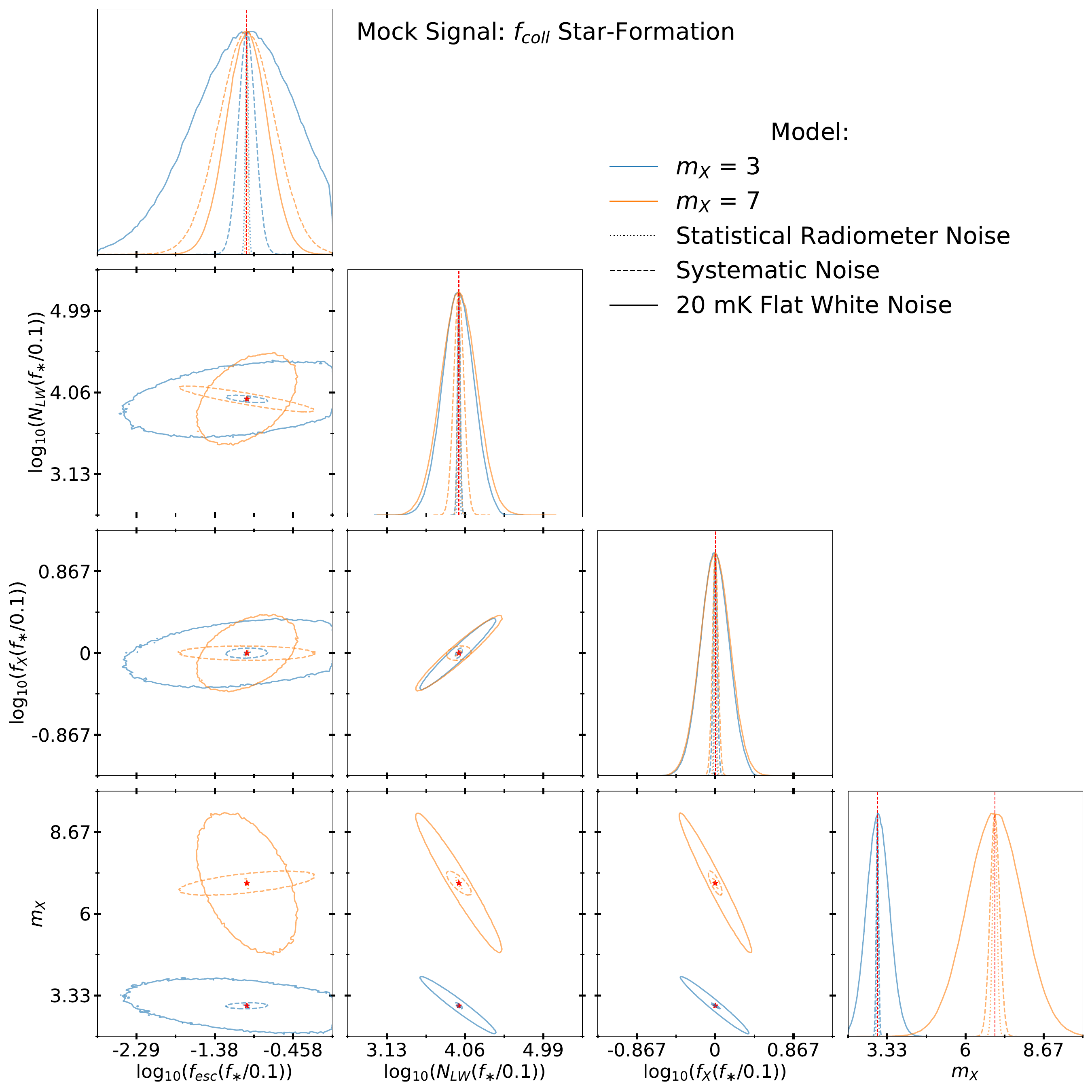}
    \caption{Triangle plot showing 95 $\%$ confidence ellipses for the collapse fraction SFR model. The three different noise levels are represented by the different linestyles. The spectrally flat 20 mK White noise is shown by solid curves, the Systematic noise by dashed, and the Statistical Radiometer noise by dotted. WDM with $m_X = 3 ,7$ keV are shown in blue and orange, respectively. The WDM mass in particular is most unconstrained for the 20 mK White noise. The degeneracies of the parameters with the star-formation efficiency $f_{\ast}$ are also explicitly labeled. The edges of the boxes represent the priors imposed upon the parameters, which is why some of the ellipses are truncated.}
    \label{fig:fcoll-triangle-plots}
\end{figure*}

Armed with models for the global signal and its uncertainties, we now present the results of our Fisher Matrix forecast. The left panel of Figure \ref{fig:mock-signals} shows the mock signals  or data vector $\vb y$ used as the input for our Fisher Matrix likelihood. The input parameter vector has the form $\boldsymbol{\theta}_{ML} = [f_{esc},N_{LW},f_X,m_X]$, and Table \ref{tab:fcoll} lists their values (marked by red stars for the 2D plots and red dashed lines for the 1D plots in Figure \ref{fig:fcoll-triangle-plots}). Additionally, we impose uniform priors upon the parameter distributions. The ranges are listed in the latter table and are further delineated by the box edges in Figure \ref{fig:fcoll-triangle-plots}, which is why many of the parameter ellipses are truncated. Finally, Figure \ref{fig:fcoll-triangle-plots} shows the 95$\%$ confidence contour ellipses from the Fisher forecasts using Equation \ref{eqn-fisher-matrix} for the collapse fraction SFR model, including all three types of input noise covariances and two different WDM masses, $m_X = 3,7$ keV.  

The dotted line shows the case for Statistical Radiometer noise, which, as expected, displays the smallest parameter uncertainties. These are then contained within the ellipses for Systematic and White noise, shown by the dashed and solid lines, respectively.

It is worth noting that $m_X = 7$ (orange contours) has uncertainties almost twice as large as the $m_X = 3$ model (blue contours), which is most apparent in the bottom right 1-D histogram of Figure \ref{fig:fcoll-triangle-plots}. The reason for this inflated uncertainty for the larger WDM mass is that the larger the WDM mass, the closer to CDM our model becomes. A greater number of small mass halos means that there are many more ways to change the astrophysical parameters, or the light produced from each halo, in order to achieve the same effect. For a slightly smaller or larger WDM mass, we still have many halos to work with to produce the same effects upon the signal, meaning that there is far greater uncertainty in what the exact DM halo mass cutoff scale should be and hence the underlying WDM mass. If on the other hand we have only a small number of low mass halos, then the rates of star formation and photon production are far more constrained, as even small changes will have a dramatic effect. Therefore, the WDM mass is better constrained at the low end. As it is the more challenging case, for the remainder of this paper we shall work with $m_X = 7$ keV in order to really test how well the global signal can contribute to constraining structure formation and DM.

\subsection{Parameter Covariance: Double-Power Law model}

\begin{figure*}
    \centering
    \includegraphics[width=0.96\textwidth]{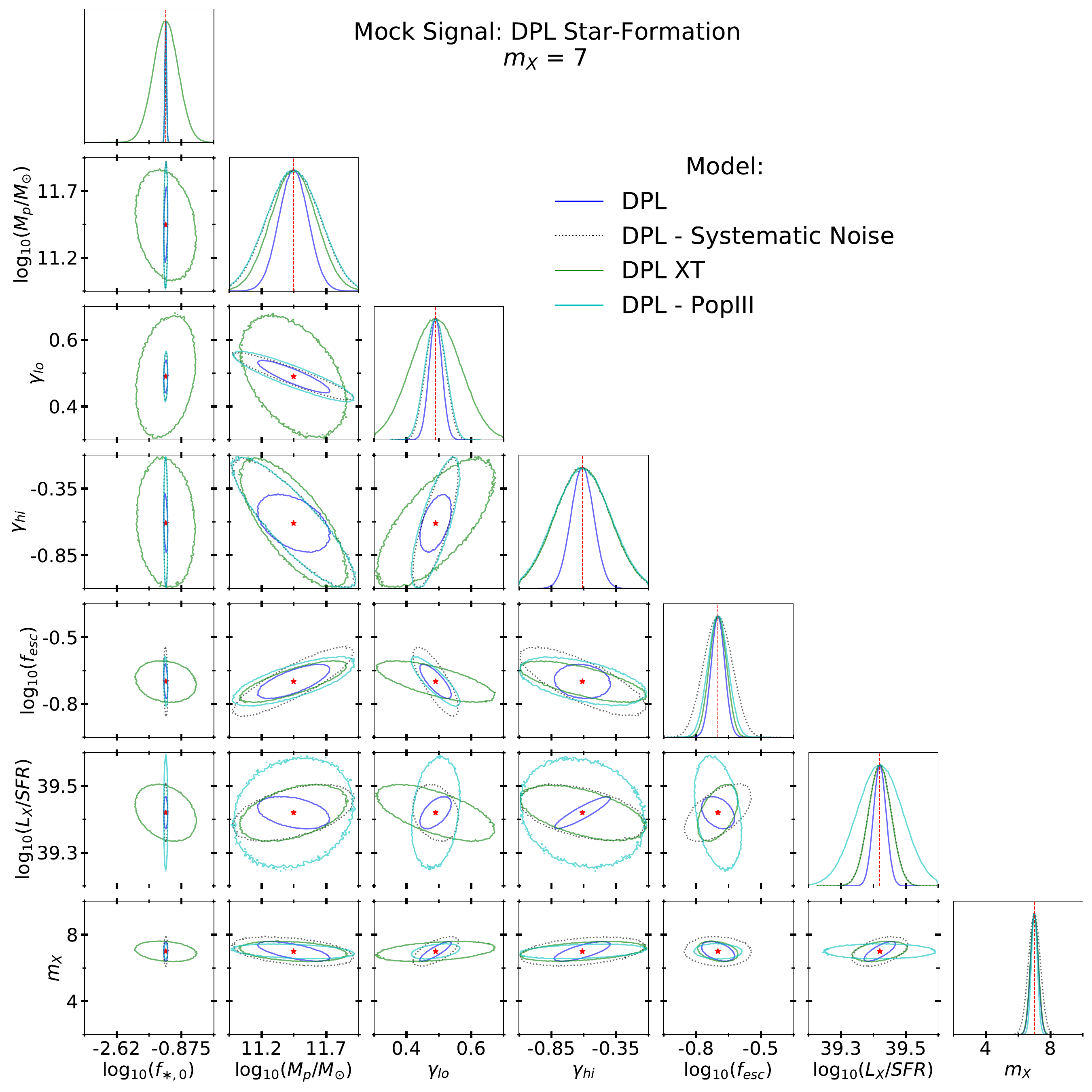}
    \caption{Triangle plot showing the 95 $\%$ confidence ellipses for a Fisher forecast using the DPL models with a DPL mock signal which does not include the effects of Pop III stars, and with $m_X = 7$ keV. Statistical Radiometer noise is used in each fit, save for the model which specifies Systematic noise (dotted black curve). Each model includes a joint fit of the global signal and UVLF, as noted in the text and in the legend. The DPL XT model (green curves) introduces nuisance parameters to explore additional star-formation effects which are potentially degenerate with WDM; these parameters are not plotted here. Similarly, the DPL - Pop III model contains our three Pop III star parameters and allows them to vary, and their covariances are also excluded from this triangle plot. The red stars and red dashed lines denote the input means (or maximum likelihood estimates) for the 2D and 1D histograms, respectively.}
    \label{fig:fisher-dpl-nopopIII}
\end{figure*}

\begin{figure*}
    \centering
    \includegraphics[width=0.96\textwidth]{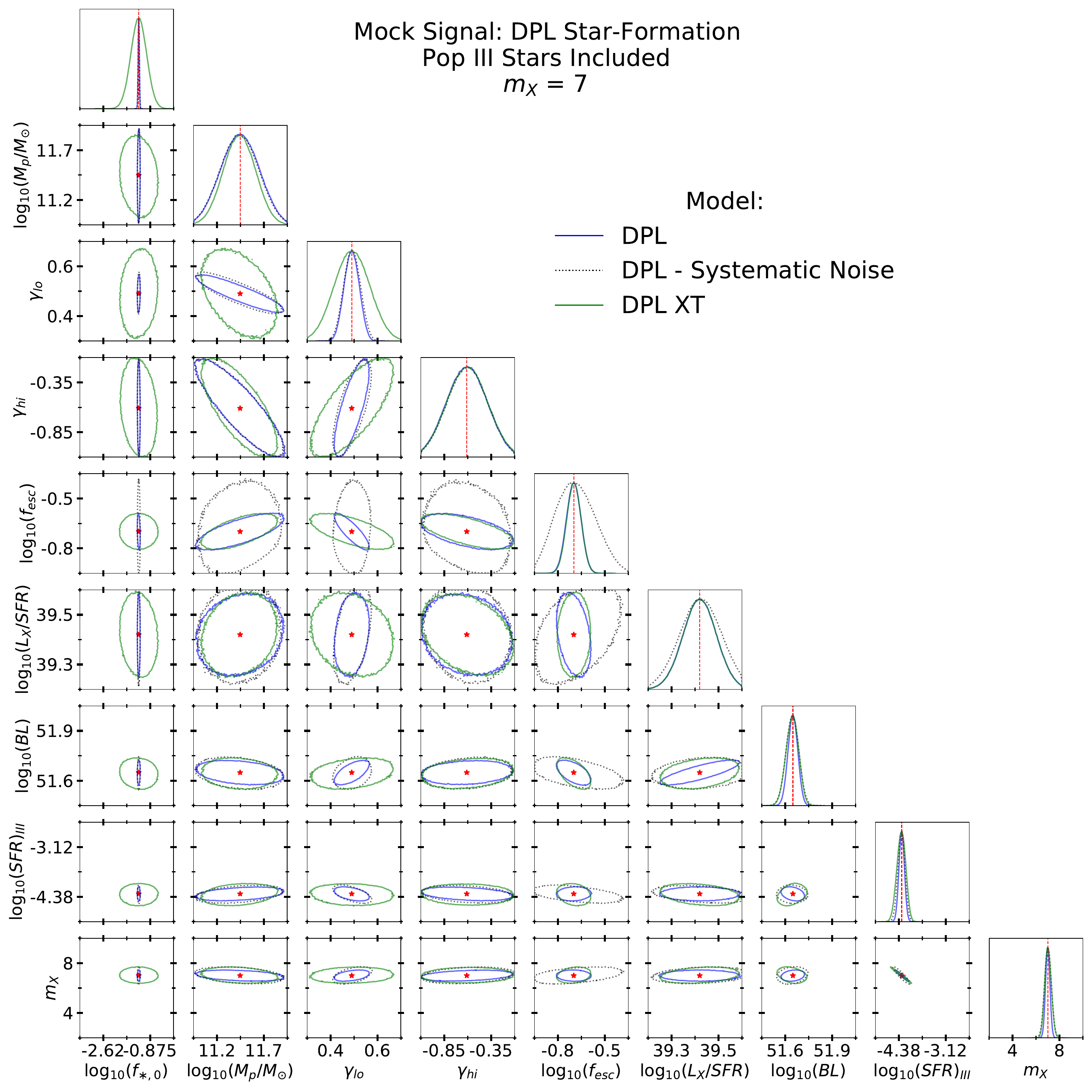}
    \caption{Triangle plot showing the 95 $\%$ confidence ellipses for a Fisher forecast using the DPL models with a DPL mock signal which includes the effects of Pop III stars, and with $m_X = 7$ keV. Statistical Radiometer noise is used in each fit, save for the model which specifies Systematic noise (dotted black curve). Each model includes a joint fit of the global signal and UVLF, as noted in the text and in the legend. The DPL XT model (green curves) introduces nuisance parameters to explore additional star-formation effects which are potentially degenerate with WDM; these parameters are not plotted here. All three of the models allow the Pop III star formation parameters to vary, although only the $SFR_{III}$ and $BL$ parameters are plotted here, as the X-ray normalization $(L_X/SFR)_{III}$ is completely unconstrained by this Fisher Matrix forecast. The red stars and red dashed lines denote the input means (or maximum likelihood estimates) for the 2D and 1D histograms, respectively.}
    \label{fig:fisher-dpl-popIII}
\end{figure*}

Figures \ref{fig:fisher-dpl-nopopIII} and \ref{fig:fisher-dpl-popIII} show the results of our Fisher forecasts respectively without and with Pop III stars included in the mock signals, which are plotted in the right panel of Figure \ref{fig:mock-signals}. For brevity we shall we refer to these DPL forecasts as the Pop II and Pop III mocks, respectively, to denote whether Pop III stars are included in the data vector. Table \ref{tab:dblpowerlaw} gives the fiducial values of the Pop II astrophysical parameters.

We purposefully choose Pop III SFE parameters for this initial forecast which only weakly affect the global signal, making it very similar to the Pop II-only case, as seen by the dotted lines denoting Pop III global signals in Figure \ref{fig:mock-signals}: if we can fit this global signal while allowing the Pop III parameters to vary and still constrain the WDM mass, then there's reason to be optimistic. Later we also consider a global signal with more efficient Pop III star formation (see Section \ref{sec-popIIIstars-methods}) for details. Tables \ref{tab:popIII} and \ref{tab:eff-popIII} show the fiducial values of the Pop III astrophysical parameters for these two cases, respectively.

We use the Statistical Radiometer noise in all cases to show the inherent parameter covariances, but also include the case of Systematic noise as a dotted black line for comparison. As in the case of the $f_{coll}$ triangle plots, we denote the mean or fiducial (MLE) values with red stars in the 2D plots and red dashed lines in the 1D histograms.

There are several points to note for the Pop II mock forecast in Figure \ref{fig:fisher-dpl-nopopIII}: first, that the DPL XT model tends to inflate the parameter covariances the most for the Statistical Radiometer noise, due to the added degeneracies between the DPL XT nuisance parameters ($a,b,c$) and the typical Pop  II parameters in the DPL model. Observe that the DPL XT model gives the poorest constraints on $f_{\ast,0}$, the SFE at peak mass, which is the best constrained by all of the other models. It is interesting, however, that this tendency for the DPL XT model to increase the parameter covariances does not extend to the WDM mass parameter, which has roughly the same magnitude of uncertainty as the model with Systematic noise at $\sim 0.5$ keV. The blue line, which represents fitting with the \textit{vanilla} DPL model used to generate the mock signal (and hence the best-fitting model) tends to have the smallest parameter covariances, as expected. In particular, $m_X$, $L_X/SFR$, $f_{esc}$\footnote{Although we haven't done so in this work, in future works we plan to include reionization constraints in the likelihood in order to help constrain $f_{esc}$.}, $\gamma_{lo}$, and $f_{\ast,0}$ are the best constrained, which is encouraging. In fact, the latter parameter is so well constrained its ellipses appear as lines in this prior range. The same is also true for the case of the DPL - Pop III model shown by the cyan curves; this model allows for variations from the three Pop III parameters, although they are not explicitly plotted in Figure \ref{fig:fisher-dpl-nopopIII}.

Moreover, the DPL - Pop III model has several interesting effects. Firstly, its Pop II astrophysical parameter covariances are nearly the same as the \textit{vanilla} DPL model (with the notable exception of the $L_X/SFR$ parameter). Secondly, the WDM mass is the best constrained of all the models, to $\sim 0.4$ keV. The UVLF is responsible for the first of these effects. Recall that the UVLF data are an estimate of the $z \sim 6$, Pop II rest-frame UV $1600 \AA$ luminosity function, and our model for it depends (primarily) upon the Pop II SFE parameters: $\boldsymbol{\mathcal{M}}_{UVLF}(f_{\ast,0}, M_p, \gamma_{lo}, \gamma_{hi}$). Varying the Pop III parameters then does not change $\boldsymbol{\mathcal{M}}_{UVLF}$, and so the DPL - P48op III model's astrophysical parameter covariances (cyan contours) come up against a wall in the parameter space imposed by the UVLF constraints, as does the model with Systematic noise (black dotted line). The fact that the DPL - Pop III model gives the strongest constraints on the WDM mass parameter is a theme that will surface again when we examine mock signals with Pop III stars included (see below).

Remarkably, the model with Systematic noise (the black dotted lines) is relatively well-behaved, despite having low-band errors nearly three orders of magnitude larger than the Statistical Radiometer noise. Indeed, in the absence of any additional nuisance or Pop III parameters, the UVLF presents a wall beyond which the SFE parameters are not allowed to vary, constraining them to the same level as in the Statistical Radiometer noise, as noted above. The $f_{esc}$ is the most poorly constrained parameter for the Systematic noise model, as the UVLF offers no help. Remarkably, the WDM mass for Systematic noise is still constrained at $7 \pm 0.7$ keV at the 95$\%$ confidence level.

Now let us turn to Figure \ref{fig:fisher-dpl-popIII}, which shows a similar Fisher forecast but when Pop III star effects are included in the mock signal (the Pop III mock) and Pop III parameters are included in all of the models. First, the DPL constraints on the SFE parameters $f_{\ast,0}, M_p, \gamma_{lo}, \gamma_{hi}$, and $f_{esc}$ are qualitatively similar to the above case where the mock signal did not include Pop III stars (blue curves in Figure \ref{fig:fisher-dpl-nopopIII}), as they are still subject to the constraining power of the UVLF. Of particular interest are the Pop III parameters. The $(L_X/SFR)_{III}$ parameter is essentially unconstrained in every model over this prior range, and is thus not plotted in Figure \ref{fig:fisher-dpl-popIII}. Evidently the values of this Pop III parameter has little effect upon the global signal at these Pop III star-formation efficiencies. Follow-up tests have shown that one needs to increase $(L_X/SFR)_{III}$ by nearly a thousand before differences in the global signal become even visually apparent.

The $SFR_{III}$ and $BL$ are parameters are both well-constrained, even for the worst case scenarios of Systematic noise or the DPL XT model. The fact the $SFR_{III}$ and $BL$ parameters are well constrained while the X-ray normalization is not must mean that our Fisher analysis is sensitive to the subtle impact on the global signal just after the onset of Wouthuysen-Field \citep{Wouthuysen:1952,Field:1959} coupling, when the Pop III stars ignite. The X-ray parameter, on the other hand, has a more cumulative impact on the global signal: it doesn't affect the initial onset of the Cosmic Dawn trough as sensitively as $SFR_{III}$ or $BL$ do. Rather, it determines the amplitude and steepness after Lyman$-\alpha$ coupling begins and the Cosmic Dawn trough appears.

For the DPL model here (blue curves), $m_X$ is constrained to the same level as the the DPL model (cyan curves) in the Pop II mock in Figure \ref{fig:fisher-dpl-nopopIII}, as most clearly seen in the 1D marginalized histogram for $m_X$ in both Figures; both give $m_X = 7 \pm 0.37$ keV. However, both the vanilla DPL model (blue) and the Systematic noise model (black dotted) in this Pop III mock become $\sim 0.37$ and $0.57$ keV, respectively, a slight improvement over the corresponding models in the Pop II mock (which had $\sim 0.48$ and $0.73$ keV, respectively).

In general, if the mock signal contains the effects of Pop III stars, then it is much easier to constrain the WDM mass, even accounting for Systematic noise and the DPL XT model. This is because adding Pop III stars not only shifts the signal to earlier times, but leaves unique signatures in the global signal, often with asymmetric features (\cite{Mirocha:2018}). Therefore, decreasing the WDM mass, leading to a later signal, cannot be necessarily compensated by, for instance, only increasing $SFR_{III}$, as was the case for the DPL - Pop III model fitting the Pop II mock. The asymmetrical features require unique combinations of Pop III parameters.

\begin{figure*}
    \centering
    \includegraphics[width=0.96\textwidth]{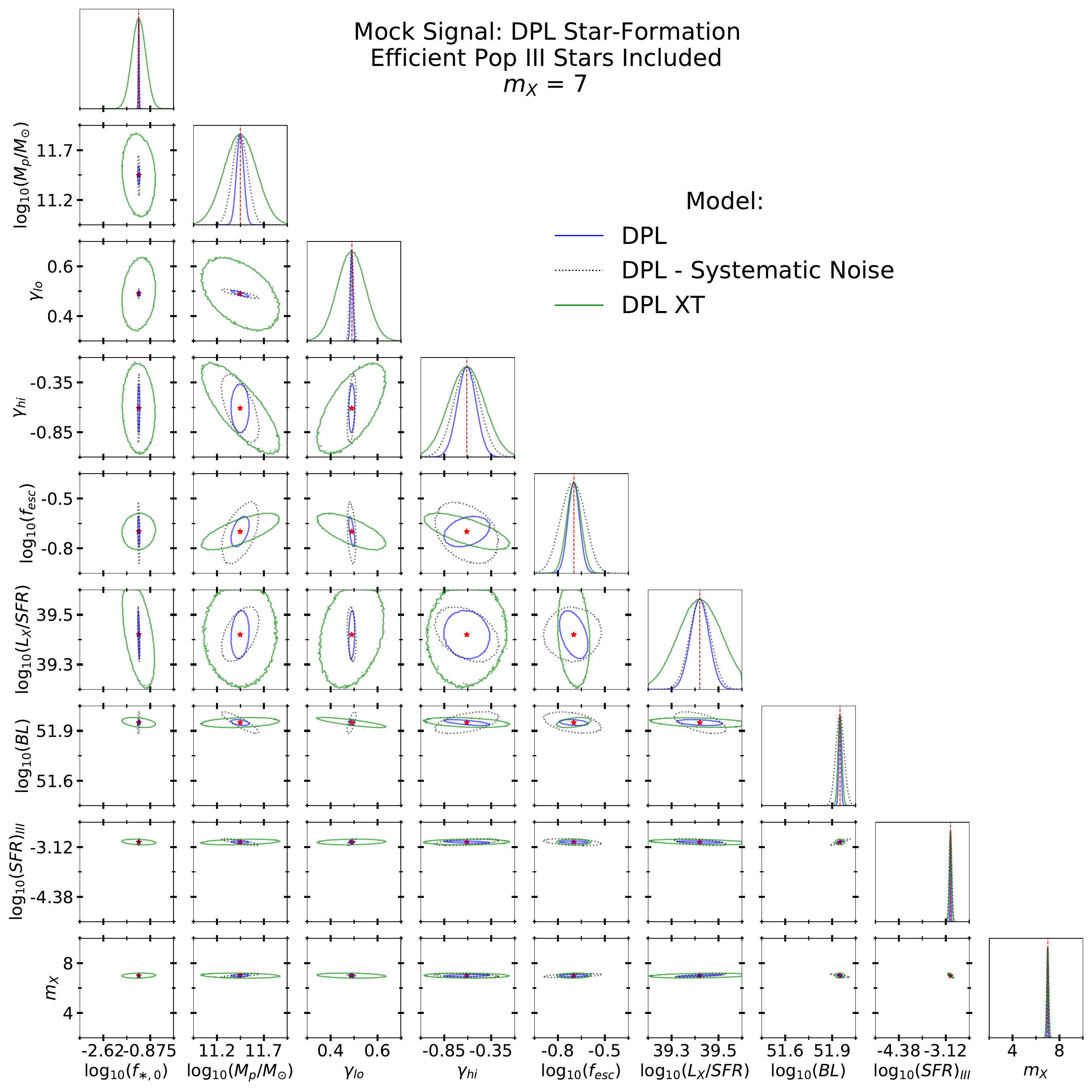}
    \caption{Same as Figure \ref{fig:fisher-dpl-popIII} but with the mock signal made using a much higher star-formation rate and binding energy limit.}
    \label{fig:app-triangleplot}
\end{figure*}

Indeed, this is most apparent when we create a mock signal with very strong Pop III effects. Figure \ref{fig:app-triangleplot} shows the same triangle plot as Figure \ref{fig:fisher-dpl-popIII} but with $SFR_{III} = 1 \times 10^{-3}$ and the binding energy limit increased to $10^{52}$ erg (shown in Table \ref{tab:eff-popIII}). All Pop III astrophysical parameters become very tightly constrained (including the X-ray normalization), and the WDM mass perhaps most of all, with WDM mass uncertainties at the 95$\%$ confidence level $\sim 0.07$ and $\sim 0.13$ keV for the Statistical Radiometer and Systematic noise, respectively. Thus, more prevalent Pop III star formation results in stronger constraints on the WDM mass.

\section{Discussion}
\label{sec-discussion}

\subsection{Feedback on Pop~III Star Formation}
Recent simulations of feedback effects on the first halos \citep{Kulkarni2021,Schauer2021} suggest a shallower relationship between the LW background intensity and minimum mass than was found in early work \citep[e.g.,][]{Machacek2001}. This implies that for a given LW background intensity, the overall impact of feedback will be weaker \citep[see also][]{Skinner2020}, giving rise to the possibility of more Pop~III star formation (as $M_{\min,\textsc{iii}}$ rises more slowly). Given that the formulae from \cite{Visbal2014} (which we employ) match the early \citet{Machacek2001} results reasonably well, we expect that adopting the newer results would increase our Pop~III SFRDs and so shift the features of the global signal slightly earlier. 

Along these lines, we have neglected the relative velocity between dark matter and baryons \citep{Tseliakhovich2010}, which is expected to suppress star formation in low-mass halos \citep{Visbal2012,OLeary2012} and so have an impact on the 21-cm signal \citep{Dalal2010,Fialkov2013,McQuinn2012}. Recent work which incorporates both effects suggests that streaming velocities set the minimum mass at $z \gtrsim 20$, with LW feedback taking over at $z \lesssim 20$ \citep[see Fig. 5 in][]{Munoz2021}. This is of course model-dependent, but shows that both effects could be important at redshifts relevant to the global 21-cm signal. 

Astrophysical assumptions alone can change the total amount of Pop~III star formation by many orders of magnitude, e.g., predictions for the peak Pop~III SFRD range between $\sim 10^{-6}$ and $\sim 10^{-3} \ M_{\odot} \ \rm{yr}^{-1} \ \rm{cMpc}^{-3}$ \citep[e.g.,][]{Mebane2018,Jaacks2018,Sarmento2019}. However, given the extreme sensitivity of the global signal to the properties of low-mass halos, we plan to address the interplay between WDM, LW feedback, and the streaming velocities in future work.

\subsection{Independent Probes}
In theory, if the Pop III stars are formed efficiently with large luminosities, then they will leave an imprint upon the UV Luminosity Function data at $z \sim 6$. Adding Pop III stars extrapolated to these lower redshifts to the UVLF data could thus help to constrain the Pop III star-formation parameters. However, in practice, our implementation assumes that a single Pop III star forms in the requisite halos, producing a spike in the UVLF at the magnitude corresponding to the minimum Pop III halo mass, which is currently beyond the data presented in \citep{Bouwens:2015a} and used in our analysis. Thus, unless Pop III star-formation is fabulously efficient, it won't have any effect upon the UVLF. In reality Pop III stars likely produce a distribution of luminosities, and the most massive stars could then affect the UVLF, but we leave such an exploration to future work.

Furthermore, independent constraints upon reionization, such as measurements of the CMB optical depth $\tau$ \citep{Planck:2020}, the neutral fraction $x_{HI}$, and the near-infrared background can also help constrain Pop III star-formation. However, recent analyses in \citet{Wu:2021} argue that the prospects for constraining Pop III star-formation via the CMB are pretty bleak. Alternatively, if the Pop III SFRD is high, $\gtrsim 10^{-3} \ M_{\odot} \ \rm{yr}^{-1} \ \rm{cMpc}^{-3}$, it is in principle detectable with SPHEREx \cite{Sun2021}. Moreover, effects on 21-cm power-spectra for similar SFRDs are non-trivial \cite{Munoz2021}, so the global signal still seems ideal.

\subsection{Astrophysical Mechanisms for the Suppression of Low-Mass Galaxies}
The other parameter which is most degenerate with the WDM mass is the minimum virial temperature assumed for star-forming halos, $T_{vir}$ or $T_{min}$. In the literature, this cutoff is sometimes modeled with a smooth exponential damping parameter applied to the HMF, such as the $M_{turn}$ parameter used in the \textsc{21cmFAST} code \citep{Mesinger:2011,Park2019}. In practice, this is simply a smoother version of using a sharp cutoff, similar to the DPL XT model employed in this study. Our Fisher forecasts make it apparent that the global signal is highly sensitive to each of these different models of suppressing low-mass galaxy formation, especially in light of the constraints still achieved on the WDM mass when implementing the DPL XT model.

As mentioned in Section \ref{sec-methodology}, we set $T_{vir}$ to a small, constant value ($T_{vir} = 300$ K, near the atomic cooling threshold) in order to study how turnovers in the HMF due to WDM alone set the scale of suppression. However, we found that even increasing $T_{vir}$ by roughly a factor of 3 (to $T_{vir} \approx 10^3$ K) for the Pop II mock models does not quantitatively change our constraints upon the WDM mass parameter (i.e. all models still give that $m_X$ is constrained to less $\lesssim 1$ keV uncertainty at $68\%$ C.I. for $m_X = 7$ keV).
We leave an exploration of the interplay between $T_{vir}$ and other mass turnover models to future work.

Although reionization feedback ought to play a similar role in characterizing the halo mass turnover, as this work focuses on $z \gtrsim 10$ this effect doesn't really matter here. So, for instance, our DPL XT is implicitly assuming then that there may be other ways to achieve a turn-over.

\subsection{The EDGES Detection and Other Experiments}
In 2018, the Experiment to Detect the Global EoR Signature (EDGES) reported the detection of an absorption trough centered at 78 MHz and with an amplitude of roughly $\sim 600$ mK -- an amplitude and shape in deep conflict with standard cosmology and CDM. Consequently, numerous explanations, from excess radio backgrounds to exotic models of DM, have been proposed to explain the signal, though it is noteworthy that no single model can explain all features of the signal (shape, amplitude, frequency). Furthermore, recent data from the SARAS global experiment were found to be inconsistent with an EDGES-like feature \citep{Singh:2021}. However, as uncertainty remains, it is possible and illustrative to use the EDGES signal to place constraints on the WDM mass. With various assumptions, several analyses have ruled out $m_X \lesssim 3$ keV \citep{Chatterjee:2019,Safarzadeh:2018}, including models for both molecular and atomic cooling halo thresholds; while \citep{Schneider:2018} found that the timing of the signal required $m_X > 6.1$ keV.

However, it should be noted that these studies only constrain WDM using the timing of the EDGES signal, and not the shape. Furthermore, they don't accurately account for degeneracies between astrophysical parameters and the WDM mass, as we have in these Fisher forecasts, and such degeneracies are certainly important for accurately determining the timing of the signal. Therefore, the research in this paper highlights the possibility that as long as the shape of the global signal is relatively well-constrained, parameter degeneracies in astrophysical models won't be fatal to inferences concerning the timing of the global signal.

With respect to the EDGES signal, the conclusions we can draw about WDM from this work are somewhat varied: on the one hand, the central redshift of the $f_{coll}$ model troughs are far more sensitive to $m_X$ and $T_{vir}$, allowing us to tune the former, for example, and see that EDGES requires $m_X \gtrsim 3$ keV from the mock signals alone; however, the amplitudes of this parametrization are four to five times shallower, from which we might conclude that a simple tuning of the WDM parameter alone in this parametrization cannot reproduce all features of the EDGES signal. On the other hand, while the DPL parametrization allows for much deeper troughs (although still not deep enough to reproduce EDGES), its central redshift is far more insouciant to changes in $m_X$ and especially $T_{vir}$, and none of the WDM examples in this work reproduce the EDGES central frequency at 78 MHz. Neither of the star-formation parametrizations reproduce the flattened Gaussian shape. And these rather rough, by-eye constraints also depend upon the assumed values of the star-formation parameters. Lastly, by including Pop III parameters and/or the DPL XT model it is certainly possible to recreate the shape and amplitude of the EDGES signal. We leave a full fit of the EDGES signal using our WDM models for future work.

While the Fisher analysis of this work using the Systematic noise assumed the observational strategies and systematics of an EDGES-like experiment, in principle any suitably characterized (in terms of beams, pointings, observation strategies, etc.) global signal experiment probing these frequency ranges will yield similar constraints upon the WDM mass parameter. Such experiments include the Shaped Antenna measurement of the background RAdio Spectrum (SARAS, \citep{Singh:2017,Singh:2021}) at 55-85 MHz; the Large-Aperture Experiment to Detect the Dark Ages (LEDA, \citep{Bernardi:2018,Garsden:2021}) at 45-83 MHz; and the Radio Experiment for the Analysis of Cosmic Hydrogen (REACH, \citep{deLeraAcedo:2019}) at 50-150 MHz.
Furthermore, interferometers such as the Hydrogen Epoch of Reionization Array (HERA, \citep{DeBoer:2017,HERA:2022}); the Square-Kilometer Array (SKA, \citep{Barry:2021}); and the Murchison Widefield Array( MWA, \citep{Li:2018}) measuring the 21 cm power spectrum can help to constrain WDM and the other star-formation parameters through a consistent, joint fit between the global signal and the power spectrum, similar to how the high-redshift UVLFs were included for the DPL models. In addition, the interferometers can constrain various global star-formation parameters separately, which can be checked against global signal models at the same redshift for consistency or even used as priors for the parameters. Such forecasts and implications for constraining non-CDM models have recently been studied in \citep{Sambit:2022} for mock data of the SKA.

\section{Conclusions}
\label{sec-conclusions}
The highly redshifted global 21-cm signal of the IGM offers a unique probe of structure formation and dark matter (DM) due to its sensitivity to the abundance of low-mass halos. We consider the case of DM with a high thermal relic velocity, known as Warm DM, characterized by its thermal mass $m_X$. In this model, low-mass halos with fluctuation wavelengths less than the free-streaming length will diffuse, resulting in a linear matter power spectrum with a smooth turnover in the low-mass halo function (see Figure \ref{fig:hmf-various-dm-models}). Small changes to the number of light-producing halos manifests in the timing and amplitude of the Cosmic Dawn trough of the global 21-cm signal (see Figure \ref{fig:global-signal}).

We examine two different high-redshift star-formation parametrizations, both of which include parameters which are potentially degenerate with the WDM mass $m_X$. The first, called the $f_{coll}$ model treats star-formation efficiency (SFE) as a constant, while the second, called the \textit{double-power law} (DPL) model, allows SFE to vary separately for the low and high mass halos. Moreover, we can use observations of high-redshift UV luminosity functions \citep{Bouwens:2015a} to further constrain its SFE parameters. We study several variants of this latter model, including one which allows for Pop III star-formation parameters and another which allows the SFE to depart from a power-law at the low-mass end.

We study the parameters of these various models using a simple Fisher matrix forecast, including several different input noise covariances for comparison: (i) Statistical Radiometer noise with a low-band error magnitude of $\sim 7$ mK at 45 MHz; (ii) Systematic noise which includes the higher uncertainties caused by the overlap between the global signal model and a beam-weighted foreground model, making the covariance non-diagonal with a low-band error magnitude of $\sim 100$ mK; (iii) 20 mK spectrally flat, diagonal White noise. 

We find that the $f_{coll}$ model's astrophysical parameters are all well constrained, with the 20 mK White noise producing the highest uncertainties. The WDM mass is poorly constrained for this noise level, however, with the case of $m_X = 7$ keV almost indistinguishable from CDM or lower WDM masses. This is because the large covariances in these astrophysical parameters allow them to compensate for virtually any WDM mass ranging from $\sim 5 - 9$ keV at the 95$\%$ confidence level. However, the case with Systematic noise is much better constrained, giving $m_X \approx 7 \pm 0.3$ keV, and the Statistical Radiometer noise is slightly better still.

Encouragingly, for the DPL model and a likelihood that contains only Pop II stars, even for models in which the SFE is steep, we can still extract $m_X = $ 7 keV WDM with high confidence. The exact uncertainties vary depending on what one assumes, ($\sim 0.5$ keV for the normal DPL model with Statistical Radiometer noise, and $\sim 0.7$ keV for Systematic noise, both at the 95$\%$ confidence level) but it's reasonable to be optimistic as, e.g. even when marginalizing over Pop III parameters we get $\sim 0.5$ keV level constraints.

When the likelihood contains Pop III stars, we find that more efficient Pop III star formation actually improves the constraints on WDM for DPL model ($\sim 0.07$ keV in the efficient Pop III star-formation case, as opposed to $\sim 0.37$ keV for low Pop III star-formation). This is because strong Pop III star-formation tends to create asymmetric features in the signal which are difficult to reproduce by simply varying the WDM mass alone. Finally, we find that the SFR of Pop III stars is the best constrained of the Pop III parameters, to within $0.4$ dex.

Overall, we are able to extract $m_X$ with reasonable uncertainties ($\lesssim 1$ keV) even when including a host of messy astrophysical parameters, models, and various, confounding noise levels. Indeed, for the fiducial value of $m_X = 7$ keV, such constraints are good enough to distinguish between WDM and CDM with high confidence. As WDM controls primarily the timing of the signal (i.e. the onset of Cosmic Dawn), this suggests that the shape of the signal is quite informative of, and sensitive to, structure formation and DM. Thus, it may be wise to consider the features of the global signal, such as the onset and slope of the Cosmic Dawn trough, as good indicators of the underlying cosmology, rather than merely parametrizing the latter trough in terms of a central frequency and amplitude, as has been done with the EDGES trough \citep{Bowman:18}.

However, if these Fisher forecasts are able to pick up on such subtle, cosmological imprints in the signal, then we have every reason to suspect that they should also pick up on similarly subtle, um-modeled instrumental or observational systematics. Such small systematic features could then change our parameter covariances dramatically. Our inclusion of Systematic noise here merely takes into account the unavoidable overlap between beam-weighted foreground and signal models, not the un-modeled component of either. Still, even accounting for such inevitable overlap it is somewhat surprising that we are able to extract reasonable uncertainties for WDM and the astrophysical parameters.

Increasing the WDM mass to $\gtrsim 10$ keV produces a DM model which is effectively indistinguishable from CDM. Because of this, we expect the posterior uncertainties on $m_X$ to become a step function, with a lower edge given by this CDM threshold. As Fisher forecasts assume all parameter posteriors are Gaussian, we expect this methodology to break down as one approaches the CDM threshold. Thus, Fisher forecast inferences above such a level become increasingly dubious, and it would be better to simply explore the likelihood using MCMC methods.

Indeed, these Fisher forecasts are preliminary work to a follow-up study to use a full MCMC algorithm to sample the large parameter spaces in question, and make a useful guide for determining which are the parameters of greatest interest, as these models are computationally and temporally expensive. In particular, including Pop III stars takes an order of magnitude longer in terms of computation time; however, this is still relatively fast compared to semi-numerical models.

\acknowledgements{The authors would like to thank Aurel Schneider for his insightful comments and ideas which helped to improve this work. This work was directly supported by the NASA (National Aeronautics and Space Administration) Solar System Exploration Virtual Institute cooperative agreement 80ARC017M0006. This work was also supported by NASA under award number NNA16BD14C for NASA Academic Mission Services.}

\bibliographystyle{yahapj}
\bibliography{references_wdm}

\end{document}